\documentclass[12pt]{article}

\pdfoutput=1
\usepackage[latin9]{inputenc}
\usepackage[top=80pt,bottom=85pt,left=67pt,right=67pt]{geometry}
\usepackage{amssymb,amsmath}
\usepackage{xypic,subfigure}
\usepackage{graphicx,color}
\usepackage{bbm}
\usepackage{float}
\usepackage{cite}
\usepackage{tensor}
\usepackage{slashed}
\usepackage[debug,pageanchor=false]{hyperref}
\definecolor{link}{rgb}{.8,.15,.1}
\hypersetup{colorlinks=true,linkcolor=link,citecolor=link,urlcolor=link,linktocpage}

 \allowdisplaybreaks

\makeatletter
\@addtoreset{equation}{section}
\makeatother

\newcommand{\beq}{\begin{equation}}
\newcommand{\eeq}{\end{equation}}
\newcommand{\bea}{\begin{eqnarray}}
\newcommand{\eea}{\end{eqnarray}}
\newcommand{\nn}{\nonumber}

\newcommand{\eq}{\begin{equation}}
\newcommand{\feq}{\end{equation}}
\newcommand{\eqn}{\begin{eqnarray}}
\newcommand{\feqn}{\end{eqnarray}}

\newcommand{\mrm}[1]{\mbox{$\mathrm{#1}$}}

\begin{document}
\begin{titlepage}

\begin{center}

\vskip .5in 
\noindent

{\Large \bf{ $\mathcal{N}=6$ supersymmetric AdS$_2 \times \mathbb{CP}^3\times \Sigma_2 $}}

\bigskip\medskip

Andrea Conti\footnote{contiandrea@uniovi.es}, Yolanda Lozano\footnote{ylozano@uniovi.es},  Niall T. Macpherson\footnote{macphersonniall@uniovi.es}  \\

\bigskip\medskip
{\small 

Department of Physics, University of Oviedo,
Avda. Calvo Sotelo s/n, 33007 Oviedo}

\medskip
{\small and}

\medskip
{\small 

Instituto Universitario de Ciencias y Tecnolog\'ias Espaciales de Asturias (ICTEA),\\
Calle de la Independencia 13, 33004 Oviedo, Spain}

\vskip 2cm 

     	{\bf Abstract }
	\end{center}
	\noindent
	
We perform a complete classification of AdS$_2$ solutions of Type II supergravity realising $\mathcal{N}=6$ supersymmetry and OSp$(6|2)$ superconformal symmetry on backgrounds that are foliations of AdS$_2 \times \mathbb{CP}^3$ over a Riemann surface $\Sigma_2$. Such solutions only exist in type IIB supergravity and are in 1 to 1 correspondence with a fourth order PDE that can be locally solved in terms of two holomorphic functions. Particular solutions in the class are the T-duals of the $\text{AdS}_3\times \mathbb{CP}^3$ and $\text{AdS}_2\times \text{S}^7$ solutions to massive Type IIA supergravity found in the literature. We discuss the field theory interpretation of the two sub-classes of solutions related to  $\text{AdS}_3\times \mathbb{CP}^3$  by Abelian and non-Abelian T-duality, which provide explicit examples for the Riemann surface an annulus or a strip. In the second case we interpret the solutions as holographic duals to baryon vertex configurations realised in D2-brane box models.
     	\noindent

\noindent

\vfill
\eject

\end{titlepage}

\tableofcontents

\section{Introduction}

Intense search for $\text{AdS}_2$ solutions of both Type II and eleven dimensional supergravities has taken place in the last years, largely motivated by the possibility of providing a microscopical description to extremal black holes by means of holography. Recent works have classified or studied ten and eleven dimensional $\text{AdS}_2$ solutions with different numbers of preserved supersymmetries \cite{DHoker:2007mci,Chiodaroli:2009yw,Chiodaroli:2009xh,Dibitetto:2018gbk,Gutperle:2018fea,Dibitetto:2018gtk,Chen:2019qib,Chen:2020mtv,Dibitetto:2019nyz,Lozano:2020txg,Lozano:2020sae,Lozano:2021rmk,Ramirez:2021tkd,Lozano:2021xxs,Lozano:2021fkk,Lozano:2022vsv,Lozano:2022swp,Conti:2023naw,Conti:2023rul,Karndumri:2024gtv}. Constructing such AdS$_2$ solutions is especially challenging, due to the high dimensionality of the internal space and the many possibilities this offers for the realisation of supersymmetry. However G-structure conditions for the preservation of minimal supersymmetry are known which can aid in this process \cite{Hong:2019wyi,Legramandi:2023fjr}.
Some of the previous settings have allowed for concrete proposals for the dual superconformal quantum mechanics, in many instances by compactification of 2d theories (see \cite{Lozano:2020txg,Lozano:2020sae,Lozano:2021rmk,Ramirez:2021tkd,Lozano:2021xxs,Lozano:2021fkk,Lozano:2022vsv,Conti:2023naw}). These $\text{AdS}_2/\text{CFT}_1$ pairs deserve further investigation, as they provide explicit, well-controlled, string theory scenarios where the black hole microscopical program can be implemented. 

Perhaps less traditionally, more recent research has focused on the interpretation of low dimensional AdS geometries, in particular $\text{AdS}_2$, as holographic duals of defect superconformal field theories. For this interpretation to hold the AdS geometry must contain a non-compact direction that in some asymptotic limit builds, together with a compact submanifold in the internal space, a higher dimensional AdS geometry. The latter is then interpreted as the holographic dual of a background CFT  in which defect degrees of freedom live, in such a way that some (lower dimensional) conformality remains, that is described holographically by  the low dimensional AdS geometry. Many examples of defect AdS/CFT duals have been constructed to date (see \cite{Arav:2020asu,Faedo:2020nol,Faedo:2020lyw,Lozano:2021fkk,Lozano:2022ouq,Lozano:2022vsv,Lozano:2022swp,Capuozzo:2024onf,Arav:2024exg,Conti:2024qgx,Conti:2024rwd,Arav:2024wyg,Faedo:2025kjf}), but there is clearly plenty of room for more detailed investigations, especially in the connection to the field theory side of the correspondence.

In this work we seek to provide a complete classification of $\text{AdS}_2\times\mathbb{CP}^3\times \Sigma_2$ solutions of type II supergravity with $\mathcal{N}=6$ supersymmetry and OSp$(6|2)$ superconformal group. Inspired by the recent work \cite{Macpherson:2023cbl}, which classified $\text{AdS}_3$ solutions of type II supergravity with $\mathcal{N}=5$ and $6$ supersymmetries, we focus in this work on the simpler $\mathcal{N}=6$ realisation, and leave  the $\mathcal{N}=5$ case for a companion follow-up paper. We also give the first steps towards the construction of the dual superconformal quantum mechanics, that should be the basis for the microscopical description of black holes with these near horizon geometries. Unfortunately,  as we will see, the field theory is realised in brane intersections preserving one dimensional $\mathcal{N}=3$ supersymmetry, which to our knowledge have not been studied so far in the literature. This forces us to take a phenomenological approach, that, we hope, will prompt further investigations of quantum mechanics with this number of supersymmetries.

The paper is structured as follows. In section \ref{construction} we present the main steps taken in the construction of the general class of solutions summarised in section \ref{IIBclass}. We begin in section \ref{review} with a review of the necessary and sufficient G-structure conditions to have a ${\cal N}=1$ AdS$_2$ solutions of type II supergravity derived in \cite{Legramandi:2023fjr}. Then in section \ref{eq:derbilins} we build 6 independent sets of spinors bi-linears that manifestly give rise to ${\cal N}=6$ solutions with SO(6) R-symmetry if they all solve the conditions of the previous section. Finally in  section \ref{solving} we sketch how we solve the necessary and sufficient conditions for supersymmetry for these spinor bi-linears. In section \ref{IIBclass} we summarise the main properties of the Type IIB solutions, especially those that will be useful for the identification of the dual superconformal quantum mechanics. This section contains two subsections in which we discuss how recent subclasses of $\text{AdS}_2\times \mathbb{CP}^3$ solutions constructed in the literature where the Riemann surface is an annulus \cite{Conti:2023rul} or a strip  \cite{Conti:2023rul,LMP} fit into our general classification. In sections \ref{ATD-CP3} and \ref{NATD-CP3} we discuss the field theory interpretation of the two subclasses of solutions related by Abelian and non-Abelian T-dualities to the $\text{AdS}_3\times \mathbb{CP}^3\times I$ solutions constructed in  \cite{Macpherson:2023cbl}. The existence of these solutions, constructed in \cite{Conti:2023rul}, prompted in fact the present search for a general classification, as it has also been the case for other classification efforts of AdS spaces in different dimensions (see for instance \cite{Lozano:2012au,Couzens:2016iot,Lozano:2019emq}). The duality relations with the solutions in  \cite{Macpherson:2023cbl} will pave the way for the construction of the brane intersections underlying the two solutions related with them by T-duality, and, thereof, for the identification of the 1d theories living in them. We will closely follow \cite{Lozano:2024idt}, where the  field theory interpretation of the $\text{AdS}_3\times \mathbb{CP}^3\times I$ solutions was investigated.  We will show that the brane intersection consists on $\mathcal{N}=3$ supersymmetric D2-brane boxes in which the D2-branes stretch between NS5-branes in two perpendicular directions, some of which forming bound states with D4-branes. We will show that the quivers associated to the solutions constructed by Abelian T-duality are identical to the ones associated to the $\text{AdS}_3\times \mathbb{CP}^3\times I$ solutions, with the only difference that the field content consists of $\mathcal{N}=4$ one dimensional multiplets instead of $(0,4)$ two dimensional ones\footnote{The quivers are in fact identical since $\mathcal{N}=4$ 1d multiplets are obtained upon dimensionally reducing $(0,4)$ 2d multiplets.}. This was expected since both solutions are related by Abelian T-duality and should therefore be dual to the same CFT. In turn, we show that the solutions related by non-Abelian T-duality have associated the same D2-brane box that underlies the Abelian T-dual solutions with additional branes that together with the F1-strings generated by the duality are interpreted as baryon vertices for the D2 and D4 branes present in the brane intersection. This interpretation seems to be a common feature (see \cite{Ramirez:2021tkd}) when the non-Abelian T-duality is performed with respect to a freely acting SL(2,$\mathbb{R}$) group  on $\text{AdS}_3$, as in the construction in \cite{Conti:2023rul}. We conclude this analysis with a discussion on the implications that holography imposes on the 1d CFTs dual to other solutions in the general class. Finally, section \ref{conclusions} contains our conclusions and a discussion of open directions for investigation. The paper is complemented with several appendices that we refer to in the main text and provide further details.

\section{Construction of the class of solutions}\label{construction}

In \cite{Legramandi:2023fjr} necessary and sufficient conditions for an AdS$_2$ solution of Type II supergravity to preserve minimal supersymmetry were derived in terms of geometric  constraints on certain spinor bilinears. In this section we leverage this result to derive a class of ${\cal N}=6$ AdS$_2$ solutions in Type IIB, while in Appendix  \ref{sec:NogoTypeIIA} we prove this class does not exist in Type IIA. The main features of the class are summarised in section \ref{IIBclass}. Uninterested readers in the details of this construction can jump directly to that section. We begin by summarising the salient parts of \cite{Legramandi:2023fjr} whose conventions we inherit.

\subsection{Necessary and sufficient G-structure conditions for ${\cal N}=1$ AdS$_2$}\label{review}
In \cite{Legramandi:2023fjr} geometric conditions for the preservation of minimal supersymmetry for AdS$_2$ solutions in type II supergravity were derived. In this section we review these before using them to construct a class of ${\cal N}=6$ solutions in the subsequent sections.

An AdS$_2$ solution in Type II supergravity is by definition decomposable in the form
\begin{align}
ds^2&=e^{2A}ds^2(\text{AdS}_2)+ds^2(\text{M}_8),\nn\\[2mm]
F &= f_{\pm} + e^{2A}\text{vol}(\text{AdS}_2)\wedge \star_8 \lambda(f_{\pm}),\nn\\[2mm]
H&=e^{2A}\text{vol}(\text{AdS}_2)\wedge H_1+ H_3,\label{eq:AdS2decompo}
\end{align}
where $F$ is the RR polyform and obeys the self duality constraint $\star\lambda(F)=F$, where when acting on a $k$-form $C_k$ we have $\lambda(C_k)=(-1)^{[\frac{k}{2}]}C_k$ (and $[\frac{k}{2}]$ means integer part of $\frac{k}{2}$). The upper/lower signs are taken in Type IIA/IIB. In the above $(e^{2A},f_{\pm},H_1,H_3)$ and the dilaton $\Phi$ have support on M$_8$, such that the SO(2,2) isometry of AdS$_2$ is preserved by the entire background. We work in conventions where the inverse radius of AdS$_2$ is $m$.

When an AdS$_2$ solution preserves supersymmetry it necessarily supports two non chiral Majorana spinors on M$_8$, namely $(\chi_1,\chi_2)$. A peculiarity of supersymmetric AdS$_2$ solutions is that they experience an enhancement to AdS$_3$ unless these spinors are constrained as
\begin{equation}
\chi^{\dag}_1\gamma_a\chi_1= \pm \chi^{\dag}_2\gamma_a\chi_2,~~~\chi^{\dag}_1\hat\gamma\chi_1= \pm \chi^{\dag}_2\hat\gamma\chi_2,~~~|\chi_1|^2=|\chi_2|^2,\label{eq:noads3}
\end{equation}
where $\gamma_a$ are a basis of flat space gamma matrices on M$_8$ and $\hat\gamma$ the corresponding chirality matrix. When these constraints are imposed supersymmetry can be expressed in terms of the following  0 and 1-form spinor bilinears
\beq
e^{A}=|\chi_1|^2,~~~e^{A}\cos\beta=\chi_1^{\dag}\hat\gamma\chi_1,~~~e^{A}\sin\beta V=\chi_1^{\dag}\gamma_a\chi_1 e^a,\label{eq:defseAVetc}
\eeq
where $e^a$ is a vielbein on M$_8$ and $V^aV_a=1$, and the following polyform bilinears
\beq
\psi=\frac{1}{16}\sum_{n=0}^8\chi^{\dag}_2\gamma_{a_n...a_1}\chi_1e^{a_1..a_n},~~~~\hat\psi=\frac{1}{16}\sum_{n=0}^8\chi^{\dag}_2\gamma_{a_n...a_1}\hat\gamma\chi_1e^{a_1..a_n},\label{eq:8d bilinearsdef}
\eeq
where it is only $(\psi_{\pm},\hat\psi_{\mp})$ that we need  in type IIA/IIB and the $\pm$ subscript indicates a projection onto the portion of these polyforms with even/odd form degree. In terms of these the conditions for supersymmetry take the form
\begin{subequations}
\begin{align}
&e^{2A}H_1=me^{A} \sin\beta V-d(e^{2A}\cos\beta),~~~~d(e^{A}\sin\beta V)=0,\label{BPS1}\\
&d_{H_3}(e^{-\Phi}\psi_{\pm})= \pm \frac{1}{16}e^{A}\sin\beta V\wedge f_{\pm},\label{BPS2}\\
&d_{H_3}(e^{A-\Phi}\hat\psi_{\mp})- m e^{-\Phi}\psi_{\pm}= \mp \frac{1}{16}e^{2A}(\star_8\lambda f_{\pm}+\cos\beta f_{\pm}),\label{BPS3}\\
&(\psi_{\pm},f_{\pm})_8=\pm \frac{1}{4} e^{-\Phi}\left(m-\frac{1}{2}e^A\sin\beta \iota_{V} H_1\right)\text{vol}(\text{M}_8),\label{BPS8dpairing}
\end{align}
\end{subequations}
where the bracket means the 8-form part of $\psi_{\pm}\wedge \lambda(f_{\pm})$.

When \eqref{BPS1}-\eqref{BPS8dpairing} hold so does supersymmetry, but this does not mean that we have a solution on shell. By definition, for that one must solve the Bianchi identities of the fluxes as well as the supergravity equations of motion. However it was shown in \cite{Legramandi:2023fjr} that, away from sources, these are all implied by 3 additional conditions
\begin{equation}
dH_3=0,~~~~\iota_V(df_{\pm} - H_3\wedge f_{\pm})=0,~~~~~d(e^{-2\Phi}\star_8H_1) + \frac{1}{2}(f_{\pm},f_{\pm})_8=0.
\end{equation}
When these also hold one has a supersymmetric solution in regular regions of the internal space M$_8$. When sources are present a little more care is required to make sure they can be identified with some physical object added to the type II action, such as a D-brane. 

\subsection{Deriving bi-linears for ${\cal N}=6$}\label{eq:derbilins}

As we seek a class of ${\cal N}=6$ solutions that support the superconformal algebra $\mathfrak{osp}(6|2)$, we need the internal space and fluxes to be compatible with an SO(6) R-symmetry. This naturally leads one to assume that the internal space contains a $\mathbb{CP}^3$. In this section and the next we focus on type IIB, we rule out the possibility of such solutions in IIA in appendix \ref{sec:NogoTypeIIA}.

$\mathbb{CP}^3$ is an example of a Kahler-Einstein manifold with SO(6) invariant Kahler form $J_2$, as such we can refine \eqref{eq:AdS2decompo} such that the metric and NS flux decompose as
\beq
ds^2(\text{M}_8)= e^{2C}ds^2(\mathbb{CP}^3)+ds^2(\text{M}_2),~~~~H_3=e^{2C}\tilde{H}_1\wedge J_2
\eeq
where now $(e^{2A},e^{2C},H_1,\tilde{H}_1,\Phi)$  are constrained to depend on the coordinates spanning M$_2$. The magnetic RR flux $f_-$ must likewise decompose in an SO(6) invariant fashion such that its dependence on $\mathbb{CP}^3$ can only come in terms of
\beq
J_2,~~~J_2\wedge J_2,~~~ \text{vol}(\mathbb{CP}^3)=\frac{1}{3!}J_2\wedge J_2 \wedge J_2,
\eeq
while its dependence on M$_2$ is to be determined. Under these assumptions the bosonic fields of the background are manifestly singlets under SO(6).

One can construct spinors and G-structure forms on $\mathbb{CP}^3$ starting from the 7-sphere which admits a parametrisation as a U(1) fibration over $\mathbb{CP}^3$ as
\beq
ds^2(\text{S}^7)= ds^2(\mathbb{CP}^3)+(e^{\tau})^2,~~~e^{\tau}=(d\tau+ \eta),~~~~ d\eta=2 J_2,\label{eq:S7metric}
\eeq
where S$^7$ is the unit radius round 7-sphere, and then dimensionally reducing on $\partial_{\tau}$. The Killing spinors on S$^7$ obey the Killing spinor equations
\beq
\nabla_{a}\xi_{\pm}=\pm \frac{i}{2} \gamma^{(7)}_a \xi_{\pm},
\eeq
where $a=1,...,7$. Naturally  $\xi_{\pm}$ transform in the \textbf{8} of SO(8) and so can in general be decomposed as
\beq
\xi_{\pm}=\sum_{I=1}^8 c^I \xi_{\pm}^I
\eeq
where the 8 independent $\xi_{\pm}^I$ can be taken to be Majorana and to obey
\beq
\xi^{I\dag}_{\pm}\xi_{\pm}^J=\delta^{IJ},
\eeq
without loss of generality and $c^I$ are 8, in general complex, constants. As we explain in greater detail in appendix \ref{S7KSE}, the $\pm$ index on these spinors has group theoretical significance. Indeed while $\xi_{\pm}^I$ are both in the \textbf{8} of SO(8) they actually furnish different representations with distinct branching rules under its SO(6) subgroup, realising respectively\footnote{Note that this is an orientation dependent statement, see appendix \ref{S7KSE} for our conventions.} 
\beq
\xi^I_+:~~~\text{\textbf{8}}\to \text{\textbf{6}}_0\oplus \text{\textbf{1}}_{-2}\oplus \text{\textbf{1}}_{2},~~~~\xi^I_-:~~~\text{\textbf{8}}\to \text{\textbf{4}}_{-1}\oplus \overline{\text{\textbf{4}}}_{1},\label{eq:branchingrules}
\eeq
where the subscript indicates the charged under the U(1) spanned by $\partial_{\tau}$. As we take them to be Majorana, each of $\xi_+^{\cal I}$ give rise to a weak G$_2$ structure with 3-form obeying
\begin{align}
\Phi^{I}_3&= -\frac{i}{3!}\xi^{({I})\dag}_+\gamma^{(7)}_{a_1a_2a_3}\xi^{({I})}_+e^{a_1a_2a_3},~~~~\star_7\Phi^{{I}}_3=-\frac{1}{4!}\xi^{{(I)}\dag}_+\gamma^{(7)}_{a_1...a_4}\xi^{{(I)}}_+e^{a_1...a_4},\nn\\[2mm]
d\Phi^{{I}}_3&=4\star_7 \Phi^{{I}}_3,~~~~~\Phi^{{I}}_3\wedge \star_7 \Phi^{{I}}_3=7\text{vol}(\text{S}^7),~~~\label{eq:weakg2forms}
\end{align}
where the brackets around an index indicate that it is not summed over. If we dimensionally reduce the 7-sphere metric in \eqref{eq:S7metric} on $\partial_{\tau}$ we are left with $ds^2(\mathbb{CP}^3)$, any portion of Killing spinors that is uncharged under this isometry survives such a reduction, clearly this is precisely the  \textbf{6}$_0$ portion of $\xi^I_+$, the remaining portion of $\xi^I_+$ are singlets under SO(6) but transforms in the \textbf{2} of the SO(2) spanned by $\partial_{\tau}$ so do not survive the reduction to $\mathbb{CP}^3$, let as denote these portions of $\xi^{I}$ as
\beq
\text{\textbf{6} of SO(6)}:~~~~ \xi_6^{\cal I},~~~~~~~~~~~~~~~~ \text{\textbf{2} of SO(2)}:~~~~ \xi_2^{\mathfrak{a}},
\eeq
where ${\cal I}=1,...,6$ and $\mathfrak{a}=1,2$. We can derive G-structure forms on $\mathbb{CP}^3$ by decomposing the weak G$_2$-structure forms into their parts parallel and orthogonal to $e^{\tau}$. In general decomposing a $d=7$ G$_2$-structure in terms of a 1-form in this fashion defines an  SU(3)-structure on the space orthogonal to it. For $I=\mathfrak{a}$ we find that 
\beq
\Phi^{a}_3= J_2\wedge e^{\tau}+\text{Im}\Omega^{\mathfrak{a}},~~~~~\star_7\Phi^{\mathfrak{a}}_3= \frac{1}{2}J_2\wedge J_2 -\text{Re}\Omega^{\mathfrak{a}}\wedge e^{\tau}
\eeq
where $(J_2,e^{\tau},\Omega^{\mathfrak{a}}_3)$ span two Sasaki-Einstein structures on the 7-sphere, \textit{i.e.}
\begin{align}
&d e^{\tau}=2J_2,~~~~~d\Omega^{\mathfrak{a}}_3=4i e^{\tau}\wedge\Omega^{\mathfrak{a}}_3,\nn\\[2mm]
&J_2\wedge J_2\wedge J_2=\frac{3}{4}i\Omega^{({\mathfrak{a}})}_3\wedge \overline{\Omega}^{({\mathfrak{a}})}_3,~~~~J_2\wedge \Omega^{\mathfrak{a}}_3=0,
\end{align}
for ${\mathfrak{a}}=1$ or $2$. Each of  $(J_2,e^{\tau},\Omega^{\mathfrak{a}}_3)$ is an SO(6) invariant by construction, but $\Omega^{({\mathfrak{a}})}_3$ is charged under the reduction isometry so it is only $J_2$ that is defined on $\mathbb{CP}^3$. Decomposing the SO(6) charged G$_2$-structure forms in a similar fashion leads to
\beq
\Phi^{{\cal I}}_3= -\tilde{J}^{{\cal I}}_2\wedge e^{\tau}+\text{Im}\Omega^{{\cal I}},~~~~~\star_7\Phi^{{\cal I}}_3= \frac{1}{2}J^{({\cal I})}_2\wedge J^{({\cal I})}_2 -\text{Re}\Omega^{\cal I}\wedge e^{\tau},
\eeq
where the 6 SU(3)-structures spanned by $(\tilde{J}^{{\cal I}}_2,\Omega^{{\cal I}}_3)$ are all charged under SO(6) and do not span the base of Sasaki-Einstein structures, rather they obey the relations
\begin{align}
& d\tilde J^{{\cal I}}_2= 4 \text{Re}\Omega^{{\cal I}}_3,~~~~d \text{Im}\Omega_3^{{\cal I}}=6 \tilde J_2^{{\cal I}}\wedge J_2-2 J_2\wedge J_2,~~~~J_2\wedge\Omega^{{\cal I}}_3=0\nn\\[2mm]
&J_2\wedge J_2+ \tilde J^{({\cal I})}_2\wedge \tilde J^{({\cal I})}_2=2 \tilde J_2\wedge J_2,~~~~J_2 \wedge J_2 \wedge J_2 = - \tilde J^{({\cal I})}_2\wedge \tilde J^{({\cal I})}_2\wedge \tilde J^{({\cal I})}_2=\text{vol}(\mathbb{CP}^3)\label{Eq:chargedrules}.
\end{align}
as well as the SU(3)-structure conditions 
\beq
\tilde{J}_2^{(\cal I)}\wedge \tilde{J}_2^{(\cal I)}\wedge \tilde{J}_2^{(\cal I)}=-\frac{3}{4}i\Omega^{(\cal I)}_3\wedge \overline{\Omega}^{(\cal I)}_3,~~~~\tilde{J}^{(\cal I)}_2\wedge\Omega^{(\cal I)}_3=0\label{Eq:SU3structurerules}.
\eeq
These forms are all singlets under the reduction isometry so define 6 independent SU(3)-structures on $\mathbb{CP}^3$ spanning the ${\cal N}=6$ supersymmetry preserved by this space.

Thus far we have identified an SO(6) sextuplet of spinors on $\mathbb{CP}^3$ $\xi_6^{\cal I}$, 6 pairs of charged forms $(\tilde{J}_2^{(\cal I)},\Omega^{(\cal I)}_3)$ spanning 6 SU(3)-structures and the SO(6) invariant $J_2$. In reducing the 7-sphere to $\mathbb{CP}^3$ the gamma matrix along the $e^{\tau}$ direction becomes a chirality matrix\footnote{We take the intertwiners defining Majorana to obey $B^{(6)}=B^{(7)}$}, \textit{i.e.} 
\beq
\gamma^{(6)}=\gamma^{(7)}_a,~~~a=1,...,6,~~~~~\hat\gamma^{(6)}=\gamma^{(7)}_7,
\eeq 
so one can define a second group of spinors in the \textbf{6} of SO(6) as $\hat\gamma^{(6)}\xi_6^{\cal I}$. It is not hard to show that this exhausts the possible options, indeed one finds that $(\xi_6^{\cal I},\hat\gamma^{(6)}\xi_6^{\cal I})$ are closed under the action of the only invariant form available to us, which is to say 
\beq
\slashed{J}_2\xi_6^{\cal I}=-i\hat\gamma^{(6)}\xi_6^{\cal I},~~~~\slashed{J}_2\hat\gamma^{(6)}\xi_6^{\cal I}=-i\xi_6^{\cal I}.
\eeq
One can show that the SO(6) spinors in $d=6$ give rise to the SO(6) charged forms as
\beq
\tilde{J}^{{\cal I}}_2=\frac{i}{2}\xi^{{(\cal I)\dag}}_6\gamma_{a_1a_2}\hat\gamma^{(6)}\xi^{{(\cal I)}}_6 e^{a_1 a_2},~~~~\text{Re}\Omega_3^{{\cal I}}=\frac{1}{3!}\xi^{{(\cal I)}\dag}_6\gamma_{a_1a_2a_3}\hat\gamma^{(6)}\xi^{{(\cal I)}}_6e^{a_1 a_2a_3},~~~~~\text{Im}\Omega_3^{{\cal I}}=-\frac{i}{3!}\xi^{{(\cal I)}\dag}_6\gamma_{a_1a_2a_3}\xi^{{(\cal I)}}_6e^{a_1 a_2a_3},\nn
\eeq
where $\xi_6^{({\cal I)}\dag}\gamma_{a_1a_2}\xi^{{{\cal I}}}_6=0$ and indices now run between $1,...,6$.

Having identified all the Majorana spinors on $\mathbb{CP}^3$ that transform in the \textbf{6} of SO(6), it is not hard to construct Majorana spinors on M$_8$ that also have this property. We decompose our $d=8$ gamma matrices, chirality matrix and intertwiner $B$ as
\begin{equation}
\begin{split}
\gamma_{a}&=\sigma_3\otimes \gamma^{(6)}_a,~~~a=1,...,6,~~~~\gamma_{7,8}=\sigma_{1,2}\otimes \mathbb{I} \\[2mm]
\hat\gamma&= \sigma_3\otimes \hat\gamma^{(6)},~~~~B=\sigma_1\otimes B^{(6)}
\end{split}
\end{equation}
for $\sigma_{1,2,3}$ the Pauli matrices and $\gamma^{(6)}_a$ the above basis of gamma matrices in $d=6$ such that $B^{(6)}=B^{(6)\dag}=(B^{(6)})^{-1}$ and $B^{(6)}\gamma^{(6)}_aB^{(6)}=-\gamma^{(6)*}_a$. Then a general sextuplet pair of spinors on M$_8$ is given by
\begin{equation}
\chi^{\cal I}_{1,2}=\frac{e^{\frac{A}{2}}}{2}\left(\eta_{1,2}\otimes\xi^{\cal I}_6 +i \hat{\eta}_{1,2}\otimes\hat{\gamma}^{(6)}\xi^{\cal I}_6\right), \\[2mm]\label{eq:sextuplets}
\end{equation}
where $(\eta_{1,2},\hat\eta_{1,2})$ are 4 independent non chiral Majorana spinors on M$_2$. 

From \eqref{eq:sextuplets} we can now construct 6  $(\psi^{\cal I}_{-},\hat\psi^{\cal I}_{+})$ pairs using \eqref{eq:8d bilinearsdef} which by construction  must be spanned by $\Omega^{{\cal I}}_3$ and wedge products of $\tilde{J}^{\cal I}_2$ and bi-linears on the 2 dimensional space orthogonal to this. However as of the 6 $(\tilde{J}^{\cal I}_2,\Omega^{{\cal I}}_3)$ pairs  obey the same relations \eqref{Eq:chargedrules}-\eqref{Eq:SU3structurerules}, and the fluxes may depend on $\mathbb{CP}^3$ though $J_2$ which comes about in the same fashion for any value of ${\cal I}$ we can safely suppress this index as
\begin{equation}
(\tilde{J}^{\cal I}_2,\Omega^{{\cal I}}_3)\to (\tilde{J}_2,\Omega_3).
\end{equation}
It is possible to show that, up to a frame rotation, the most general set of 4 spinors on M$_2$ that is consistent with \eqref{eq:noads3}, \eqref{eq:defseAVetc} and \eqref{BPS3} (specifically that $f_-$ is an SO(6) singlet) are given by
\begin{align}
\eta_1&=i\left(\begin{array}{c}
 \cos\left(\frac{\alpha-\beta}{2}\right)+i \sin\left(\frac{\alpha+\beta}{2}\right)\\-\left(\cos\left(\frac{\alpha-\beta}{2}\right)-i \sin\left(\frac{\alpha+\beta}{2}\right)\right)\end{array}\right),~~~~\hat\eta_1=-\left(\begin{array}{c}
 \cos\left(\frac{\alpha+\beta}{2}\right)+i \sin\left(\frac{\alpha-\beta}{2}\right)\\\cos\left(\frac{\alpha+\beta}{2}\right)-i \sin\left(\frac{\alpha-\beta}{2}\right)\end{array}\right),\nn\\[2mm]
\eta_2&=-i\left(\begin{array}{c}
 \cos\left(\frac{\alpha+\beta}{2}\right)+i \cos\left(\frac{\alpha-\beta}{2}\right)\\-\left(\cos\left(\frac{\alpha+\beta}{2}\right)-i \sin\left(\frac{\alpha-\beta}{2}\right)\right)\end{array}\right),~~~~\hat\eta_2=-\left(\begin{array}{c}
 \cos\left(\frac{\alpha-\beta}{2}\right)+i \sin\left(\frac{\alpha+\beta}{2}\right)\\\cos\left(\frac{\alpha-\beta}{2}\right)-i \sin\left(\frac{\alpha+\beta}{2}\right)\end{array}\right),
\end{align}
where $(\alpha,\beta)$ are functions on M$_2$ to be determined, the later being the same $\beta$ appearing in \eqref{BPS1}-\eqref{BPS8dpairing}. These yield the following decomposition of the $d=8$ bilinears in Type IIB
\begin{align}
\hat \psi_+&= -\frac{e^{A}\sin\beta}{16}\text{Re}\psi^{(7)}_+, ~~~~\psi_-=\frac{e^A}{16}\text{Re}\bigg(\psi^{(7)}_-+\cos\beta \psi^{(7)}_+\wedge V \bigg), \nn \\[2mm]
\psi^{(7)}_+ & =-e^{-i \alpha}e^{i e^{2C}\tilde{J}_2}-e^{3C}i \Omega_3\wedge U,~~~~\psi^{(7)}_-=-e^{-i \alpha}e^{i e^{2C}\tilde{J}_2}\wedge U-e^{3C}i \Omega_3,\label{eq:bilinears}
\end{align}
where $\text{vol}(\Sigma_2)=U\wedge V$. 

With $d=8$ bilinears in hand we set about solving the conditions \eqref{BPS1}-\eqref{BPS8dpairing} in the next section under the understanding that while generically these only imply ${\cal N}=1$ supersymmetry, we are actually solving 6 independent ${\cal N}=1$ sub-sectors simultaneously due to the suppressed index ${\cal I}$ which yield the claimed ${\cal N}=6$.

\subsection{Solving the supersymmetry constraints} \label{solving}
In this subsection we sketch how we solve the necessary and sufficient conditions for supersymmetry following from the bilinears derived in the previous subsection.\\
~\\
Upon substituting \eqref{eq:bilinears} into \eqref{BPS1}-\eqref{BPS8dpairing} and making use of \eqref{Eq:SU3structurerules} one finds a large and highly redundant system of equations on $\Sigma_2$ that imply supersymmetry. This system decomposes into parts that appear wedged with SO(6) singlets on $\mathbb{CP}^3$, and can contribute to the fluxes, and charged parts that appear wedged with one of the charged forms
\beq
\tilde{J}_2,~~~~J_2\wedge \tilde{J}_2,~~~~\text{Re}\Omega_3,~~~~\text{Im}\Omega_3
\eeq 
and so cannot contribute to flux. It is convenient to deal with the charged portions of \eqref{BPS2} and \eqref{BPS3} as well as the second condition in \eqref{BPS1} first. After some effort to simplify the constraints and eliminate redundancies one arrives at the following conditions
\begin{subequations}
\begin{align}
&e^C=-\frac{4 e^{A}}{ \sin\alpha \sin\beta}\, ,~~~~m e^{2C}\tilde{H}=4d(e^{A+C}\cos\alpha \sin\beta)+8 e^{A}\sin\beta U\label{eq:chargedcond1}\\[2mm]
&d(e^{A}\sin\beta U)=0,~~~~d(e^{A}\sin\beta V)=0,\label{eq:chargedcond2}\\[2mm]
&d(e^{A+3C-\Phi})+4 e^{A+2C-\Phi}(\cos\alpha U-\sin\alpha \cos\beta V)=0\label{eq:chargedcond3}.
\end{align}
\end{subequations}
The first of these conditions furnishes us with a definition of $e^{C}$ that implies both $\sin\alpha \neq 0,~\sin\beta \neq 0$ and defines the magnetic part of the NS flux which, by \eqref{eq:chargedcond2}, is necessarily closed. We can take \eqref{eq:chargedcond2} to define a vielbein on $\Sigma_2$ by introducing local coordinates $(x_1,x_2)$ such that
\beq
U+i V=  \frac{p}{e^{A}\sin\beta m}(dx_1+i dx_2),
\eeq
for $p$ a constant. Next we redefine the dilaton in terms of an arbitrary function $h=h(x_1,x_2)$  as
\beq
e^{-\Phi}= e^{-(A+3C)} h,
\eeq
so that \eqref{eq:chargedcond3} now reduces to 2 PDEs
\beq
\frac{m p \cos\beta}{e^{2A}\sin^2\beta}+\partial_{x_1}\log h=0,~~~~\frac{m p \cos\alpha}{e^{2A}\sin\alpha \sin^2\beta}+\partial_{x_2}\log h=0,\label{eq:secondcond}
\eeq
which when $\cos\beta\neq 0$ and $\cos\alpha\neq 0$ can be taken to define $\alpha$ and $e^{2A}$. One should of course be more careful when $\cos\beta= 0$ or $\cos\alpha= 0$, in which case $h$ becomes independent of one of $x_1$ or $x_2$ - however the final result one obtains by solving  \eqref{eq:secondcond} under the assumption that $\cos\beta\neq 0$ and $\cos\alpha\neq 0$ ends up being consistent with what one derives after fixing one of $\cos\beta$ or $\cos\alpha$ to zero anyway (setting both to zero is not possible).

Having solved the charged parts of \eqref{BPS2} and \eqref{BPS3} we can now extract $f_-$ from the singlet part of \eqref{BPS3} and insert the result into what remains of \eqref{BPS2}, which we find is implied, and \eqref{BPS8dpairing} which yields a final constraint
\beq
\cos\beta=-\frac{\partial_{x_1}h}{2 h(\partial_{x_1}^2+\partial_{x_2}^2)h-\partial_{x_2}h^2}.
\eeq
Finally one can substitute what has been derived thus far into the first expression in \eqref{BPS1} to extract the electric part of the NS flux. At this point the supersymmetry constraints are completely solved. We present the resulting class in the next section, where we have fixed
\beq
 p=\frac{L^2}{m},
\eeq
and use the freedom to rescale the string coupling $g_s$ (which in our conventions is 1) to send
\beq
(e^{-\Phi},F_-)\to c (e^{-\Phi},F_-).
\eeq

\section{The class of $\text{AdS}_2\times \mathbb{CP}^3\times \Sigma_2$ solutions } \label{IIBclass}
In the previous section we constructed a general class of $\mathcal{N}=6$ supersymmetric solutions in Type IIB supergravity with OSp$(6|2)$ superconformal group, here we summerise the results of that construction. The NS sector for the class is given by
\begin{align}
\frac{ds^2}{L^2}&= \sqrt{\Delta_1}\bigg[\frac{h}{\Delta_2}ds^2(\text{AdS}_2)+ \frac{1}{m^2}\bigg(\frac{8}{ \square h}ds^2(\mathbb{CP}^3)+ \frac{1}{ h}(dx_i)^2\bigg)\bigg],~~~~e^{-\Phi}=\frac{m^3 c \sqrt{h\Delta_2}(\square h)^{\frac{3}{2}}}{16\sqrt{2}L^4\Delta_1}, \nn \\[2mm]
H&=dB,~~~~\frac{B}{L^2}=\left(\frac{h \partial_1 h}{\Delta_2}+x_1\right)\text{vol}(\text{AdS}_2)+ \frac{8}{m^2}\left(\frac{\partial_2 h}{\square h}-(x_2-l)\right)J_2, \label{NSNSIIBclass}
\end{align}
where $J_2$ is the closed SO(6) invariant K\"ahler form on $\mathbb{CP}^3$ and $l$ is an integration constant/large gauge transformation whose role will be better explained when we turn into the field theory analysis in sections \ref{ATD-CP3} and \ref{NATD-CP3}. We use the notation 
\beq \label{PDE}
\square h= (\partial_{x_1}^2+\partial_{x_2}^2)h,~~~~ \partial_i= \partial_{x_i},
\eeq
and 
\beq
\Delta_1=2 h\square h-(\partial_2 h)^2 ,~~~~\Delta_2=2 h\square h-(\partial_1 h)^2 -(\partial_2 h)^2\, .
\eeq

The RR fluxes are a little esoteric. The easiest thing to present is the magnetic Page fluxes. That is to say the RR flux decomposes into magnetic and electric parts as
\beq
F_-= f_-+ e^{2A}\text{vol}(\text{AdS}_2)\wedge \star_8\lambda(f_-),
\eeq
with the full Page fluxes, given by $\hat F_-= e^{-B}\wedge F_-$, similarly decomposing as
\beq
\hat{F}_-= \hat{f}_-+ \text{vol}(\text{AdS}_2)\wedge \hat{g}_-,\label{eq:pagefluxdecomp}
\eeq
where a solution requires $d\hat{F}_-=0$ away from sources.
The magnetic Page fluxes take the form
\begin{align}
\hat f_1&=F_1= \frac{m^3 c}{32L^4}\bigg[\partial_2(\square h) dx_1-\partial_1(\square h) dx_2-d\left(\frac{\partial_1h\partial_2h\square h}{\Delta_1}\right)\bigg],\nn\\[2mm]
\hat f_3&= \frac{m c}{4L^2}\bigg[\bigg((x_2-l)\partial_2 (\square h)-\square h\bigg)dx_1-(x_2-l)\partial_1 (\square h)dx_2+d\left(\frac{\partial_1 h \square h(2 h-(x_2-l)\partial_2 h)}{\Delta_1}\right)\bigg]\wedge J_2,\nn\\[2mm]
\hat f_5&= \frac{c}{m}\bigg[\left(2 \partial_2h+(x_2-l)\left(-2 \square h+(x_2-l)\partial_2 (\square h)\right)\right)dx_1-\bigg(2 \partial_1 h+(x_2-l)^2 \partial_1(\square h)\bigg)dx_2\nn\\[2mm]
&-d\left(\frac{\partial_1h\left((x_2-l)^2 \partial_2 h\square h+2 h\left(\partial_2 h-2(x_2-l) \square h\right)\right)}{\Delta_1}\right)\bigg]\wedge J_2\wedge J_2,\nn\\[2mm]
\hat f_7&= \frac{8L^2 c}{m^3}\bigg[\left(-2 h+(x_2-l)\left(2\partial_2h+(x_2-l)\left(\frac{1}{3}(x_2-l)\partial_2(\square h)-\square h\right)\right)\right)dx_1\nn\\[2mm]
&-2\bigg((x_2-l)\partial_1 h+\frac{1}{6}\partial_1(\square h)\bigg)dx_2\nn\\[2mm]
&+d\left(\frac{\partial_1 h\left(4 h^2-(x_2-l)^3\partial_2 h \square h-6(x_2-l)h\left(\partial_2 h-(x_2-l)\square h\right)\right)}{3\Delta_1}\right)\bigg]\wedge J_2\wedge J_2\wedge J_2. \label{RRIIBclass}
\end{align}
The electric ones, that will be needed in the field theory analysis that we will perform in the following sections, are collected in appendix \ref{Sec:electricpagefluxes}.

We find that
\begin{align}
d\hat f_{2n+1}\sim(x_2-l)^n \square^2 h dx_1\wedge dx_2\wedge J_2^n
\end{align}
for $n=0,1,2,3$, so away from sources $h$ obeys
\beq \label{eq:PDE}
\square^2 h= (\partial_1^4+2\partial_1^2\partial_2^2+\partial_2^4)h=0,
\eeq
which is the only PDE governing the system. As the PDE is not a supersymmetry condition but rather follows from the Bianchi identities of the fluxes we can also add source terms to the PDE, but some care is required to make sure these make sense. We note that 
\beq
d\hat g_{2n+1}\sim x_1(x_2-l)^n\square^2 h dx_1\wedge dx_2\wedge J_2^n\label{eq: bianchielectric}
\eeq
for $n=0,1,2,3$. Thus if
\beq
\square^2 h\sim \delta(x_2-l)\delta(x_1)
\eeq
then only $dF_1$ receives a source term, the others vanishing as $x \delta(x)\to 0$. We further note that if a solution can be found for which $\square h$ diverges like a D7 at a given point where  $(h, \partial_i h)$ are finite then the metric and dilaton will have D7 like behaviour.  This suggests that it may be possible to construct global solutions with a boundary at $x_1=0$ along which source D7 branes can be placed at the points $x_2=l\in \mathbb{Z}$. We will explore such possibilities as well as well as a generically ${\cal N}=5$ preserving generalisation of this class in \cite{CLM}.

In the next subsection we find the general local solution to the defining PDE, given by equation \eqref{eq:PDE}. In the following subsections we will show how some particular solutions constructed recently in the literature embed in this class. Some of these solutions will be the basis of our field theory discussion in sections \ref{ATD-CP3} and \ref{NATD-CP3}.

\subsection{General local solution in terms of harmonic functions}
Away from possible sources the defining PDE \eqref{eq:PDE} can be written in terms of a complex coordinate and derivatives
\beq
z=x_1+i x_2,~~~~\partial_z=\frac{1}{2}(\partial_1-i \partial_2),~~~~\partial_{\bar{z}}=\frac{1}{2}(\partial_1+ i \partial_2),~~~~\Box=4\partial_z\partial_{\bar{z}}
\eeq
as
\beq
\partial_z\partial_{\bar{z}}\partial_z\partial_{\bar{z}}h=0
\eeq
Given that $\square h$ must be real,  we can integrate this in the equivalent form
\beq
\partial_z\partial_{\bar{z}}h=  \frac{1}{2}\left(\partial_z{\cal A}_1(z)+\partial_{\bar{z}}\bar{\cal A}_1(\bar{z})\right)
\eeq
where ${\cal A}_1(z)$ is any holomorphic function. This is a simple PDE consisting of a homogeneous and inhomogeneous part. Its general solution is given in terms of a second holomorphic function ${\cal A}_2(z)$ as
\beq
h= \frac{1}{2}\left(\bar{z}{\cal A}_1(z)+z\bar{{\cal A}}_1(\bar{z})+ {\cal A}_2(z)+ \bar{{\cal A}}_2(\bar{z})\right).
\eeq
Decomposing our two holomorphic functions in term of harmonic functions $(h_1,h_2)$ of $(x_1,x_2)$ and their harmonic duals $(h^D_1,h^D_2)$ as
\beq
{\cal A}_{1,2}= h_{1,2}+i h_{1,2}^D,
\eeq
leads to 
\beq
h=  x_1 h_1 +x_2 h_1^D+ h_2
\eeq

In the next subsections we will analyse some particular solutions to the defining PDE, first for the Riemann surface an annulus and then for a strip.

\subsection{ AdS$_2 \times \mathbb{CP}^3 \times$ S$^1 \times I$}\label{annulus}

We consider first the case in which $\Sigma_2$ is an annulus. This is the simplest case that solves the Bianchi identity \eqref{eq:PDE}, with one of the two directions spanning the Riemann surface an isometry and $h$ a function of the other coordinate.

\vspace{0.2cm}

\noindent {\bf First class:}

We start discussing the embedding of one of the two classes of $\mathcal{N}=6$ solutions derived in \cite{Conti:2023rul}. This class was obtained T-dualising the $\text{AdS}_3\times \mathbb{CP}^3\times I$ solutions in \cite{Macpherson:2023cbl} along the $\text{S}^1$ fibre of the $\text{AdS}_3$ subspace, in its parametrisation 
\begin{equation}\label{param}
ds^2(\text{AdS}_3) =\frac14\Bigl[\left( 2 dx_1+\eta \right)^2+ds^2(\text{AdS}_2)\Bigr] \qquad \text{with} \qquad d\eta=\text{vol}(\text{AdS}_2),
\end{equation}
giving rise to a $\text{AdS}_2\times \mathbb{CP}^3\times \Sigma_2$ solution of Type IIB with $\Sigma_2$ an annulus. Note that here we have taken AdS$_2$ to have unit radius, \textit{i.e.}
\beq
m=1.
\eeq
 We will discuss the field theory interpretation of this class of solutions in the next section. We can easily see that they arise from the general class \eqref{NSNSIIBclass}, \eqref{RRIIBclass} imposing that $\partial_{x_1}$ is an isometry, such that the function $h$  depends only on $x_2$, $h= h(x_2)$. In this way $h$ must be  a third order polynomial function in $x_2$, \textit{i.e.} $h^{(4)}=0$ in order to satisfy the PDE \eqref{eq:PDE}, which holds away from sources. The solution becomes
\begin{equation}
\begin{split}
\frac{ds^2}{L^2}& = \frac{h}{\sqrt{\Delta}} ds^2(\text{AdS}_2)+ \sqrt{\Delta}\bigg(\frac{8}{h''} ds^2(\mathbb{CP}^3)+ \frac{1}{h}(dx_i)^2\bigg) , ~~~~e^{-\Phi}= \frac{c \sqrt{h }(h'')^{\frac{3}{2}}}{ 16\sqrt{2}L^4 \sqrt{\Delta}},\label{andrea-class1} \\[2mm]
H&=dB,~~~~\frac{B}{L^2}= x_1 \text{vol}(\text{AdS}_2)+ 8 \left(\frac{h'}{h''}-(x_2-l)\right)J_2, \qquad \Delta = 2 h h''-h'^2,
\end{split}
\end{equation}
where AdS$_2$ has unit radius. The magnetic Page fluxes are
\begin{align}
\hat f_1& = F_1 = \frac{c}{32 L^4} h''' dx_1 ,\qquad \hat f_3 = \frac{c}{4L^2} \left( (x_2-l) h''' - h'' \right) dx_1 \wedge J_2,\nn\\[2mm]
\hat f_5& = c \left( 2 h' + (x_2-l) \left(-2 h'' + (x_2-l)h''' \right)\right)dx_1\wedge J_2\wedge J_2 \label{andrea-class1RR} .
\end{align}
One can easily check that this is the class of AdS$_2 \times \mathbb{CP}^3 \times$ S$^1 \times I$ solutions derived in \cite{Conti:2023rul}, that we summarise in the next section, with
\begin{equation}
x_1 \to r, \qquad x_2 \to \frac{1}{\pi} y, \qquad L^2 \to \frac{\pi}{2}, \qquad c \to 4 \pi^2.
\end{equation}

\vspace{0.2cm}

\noindent {\bf Second class:}

A second class of $\text{AdS}_2\times \mathbb{CP}^3\times \Sigma_2$ solutions with $\Sigma_2$ an annulus is obtained upon T-dualising the class of $\text{AdS}_2\times\text{S}^7\times I$ solutions to massive Type IIA supergravity constructed in \cite{Dibitetto:2018gbk}. This class was generalised in \cite{LMP} such that the $\text{S}^7$ is replaced by $\text{S}^7/\mathbb{Z}_k$. The T-duality takes place along the $\text{S}^1$ fibre of the $ \text{S}^7/\mathbb{Z}_k$ space, in its parametrisation
\begin{equation}\label{tau-direction}
ds^2(\text{S}^7/\mathbb{Z}_k) = \left(\frac{d\tau}{k}+{\cal A}\right)^2+ds^2(\mathbb{CP}^3).
\end{equation}
This class of solutions in Type IIB supergravity was constructed in \cite{LMP}, and was shown to be related by S-duality  to the class \eqref{andrea-class1}, \eqref{andrea-class1RR}
for vanishing axion field. One can easily check that it is obtained from the solutions in section 3 imposing that $\partial_{x_2}$ is an isometry, such that $h=h(x_1)$. The solution becomes
\begin{align}\label{SdualAbelian}
\frac{ds^2}{L^2}&= \frac{\sqrt{2}h^{3/2} \sqrt{h''}}{\Delta } ds^2(\text{AdS}_2) + \sqrt{2} \left(\frac{8 \sqrt{h}}{ \sqrt{h''}} \, ds^2(\mathbb{CP}^3) + \frac{ \sqrt{h''}}{\sqrt{h}}(dx_i)^2\right),~~~~ e^{-\Phi}=\frac{c \sqrt{h''}\sqrt{\Delta}}{32 \sqrt{2} L^4 \sqrt{h}} ,\notag \\[2mm]
H&=dB,~~~~\frac{B}{L^2}=\left(\frac{h h'}{\Delta} + x_1 \right)\text{vol}(\text{AdS}_2) - 8 (x_2-l)J_2, \qquad \Delta = 2 hh''-h'^2,
\end{align}
where again we took AdS$_2$ to have unit radius. The magnetic Page fluxes are
\begin{align}
\hat f_1& = F_1 = \frac{c }{32 L^4} h''' dx_1, \qquad \hat f_3 = - \frac{c}{4L^2} (x_2-l) h''' \, dx_2 \wedge J_2, \nn \\[2mm]
\hat f_5& = c \left( d \left( 2 (x_2 - l) h' \right) -2 (x_2-l) h'' dx_1 - \left( 2 h+ h''' (x_2-l)^2 \right) dx_2 \right) J_2 \wedge J_2, \nn \\[2mm]
\hat f_7 &  = 8 L^2 c  \left(dx_1 \left(-(x_2-l)^2 h''-2 h \right)-\frac{1}{3} dx_2 (x_2-l) \bigg(h''' (x_2-l)^2+6 h' \right) \nn \\[2mm]
& +d\left(\frac{h' \left(3 (l-x_2)^2 h''+2 h\right)}{3 h''}\right)\bigg).
\end{align}
In particular, if we redefine 
\begin{equation}
L \to \frac{L}{2 \, 2^{3/4}}, \qquad c \to \frac{L^2 }{2 \sqrt{2} \, c_0 \, k}, \qquad x_1 \to -r, \qquad x_2 \to \frac{ 2 \, \sqrt{2} \, k}{L^2}\tau,
\end{equation}
in \eqref{SdualAbelian} we obtain \cite{LMP}. \\[2mm]
The CFT dual to this class of solutions was discussed in \cite{LMP}. In that reference it was shown that the underlying brane intersection consists on M2-branes probing a $\mathbb{C}^4/\mathbb{Z}_k$ singularity with momentum, that once reduced to Type IIA and T-dualised to Type IIB admit extra D7-branes that do not break any further supersymmetries. Explicit quiver constructions were provided that allowed to interpret the solutions as dual to backreacted baryon vertices, similar to what we will encounter in our discussion in section \ref{NATD-CP3}. 

\subsection{ AdS$_2 \times \mathbb{CP}^3 \times I \times I$}\label{strip}

The second simplest case that solves the Bianchi identity \eqref{eq:PDE} is when $h$ is the product of a linear function on one of the two coordinates $x_1$ or $x_2$ and  a cubic polynomial $\hat{h}$ on the other coordinate. One can easily check that this solves  \eqref{eq:PDE} with $\hat{h}''''=0$.

Interestingly, the first possibility
\begin{equation}
\hat{h}(x_2), \label{eq:NATDmap1}
\end{equation}
gives rise to the second class of $\text{AdS}_2\times \mathbb{CP}^3$ solutions in Type IIB derived in \cite{Conti:2023rul}, where $\Sigma_2$ is a strip. This class of solutions was obtained from the $\text{AdS}_3\times \mathbb{CP}^3\times I$ solutions in \cite{Macpherson:2023cbl} by a non-Abelian T-duality transformation with respect to the left-action of the SL(2,$\mathbb{R}$) isometry group of $\text{AdS}_3$. It will be further  discussed  in section \ref{NATD-CP3}, where we will focus on its field theory interpretation. The class of solutions is given by 
\begin{align}\label{andrea-class2}
\frac{ds^2}{L^2} & = \frac{x_1^2 \hat{h} \sqrt{\Delta} }{x_1^2 \Delta - \hat{h}^2} ds^2(\text{AdS}_2)+ \frac{8 \sqrt{\Delta}}{\hat{h}''} ds^2(\mathbb{CP}^3) + \frac{ \sqrt{\Delta}}{\hat{h}} (dx_i^2), \nn \\[2mm] 
e^{-\Phi} &= \frac{c \sqrt{\hat{h}} \hat{h}''^{3/2}\sqrt{x_1^2 \Delta - \hat{h}^2} }{16 \sqrt{2} L^4 \Delta}, \qquad \Delta  = -\hat{h}'^2+2 \hat{h} \hat{h}'' , \\[2mm]
H & = dB, \qquad \frac{B}{L^2} = x_1 \left( 1 + \frac{\hat{h}^2}{x_1^2 \Delta -\hat{h}^2 }\right) \text{vol}(\text{AdS}_2)+ 8 \left( \frac{\hat{h}'}{\hat{h}''} -(x_2-l)\right)\wedge J_2. \nn
\end{align}
As we have anticipated previously, AdS$_2$ is the unit radius metric on the two dimensional Anti-de-Sitter spacetime. The magnetic RR Page fluxes are
\begin{align}
\hat{f}_1 & = \frac{c}{L^4} \left( \hat{h}''' x_1 dx_1 - \hat{h}'' dx_2 - d \left( \frac{\hat{h} \hat{h}' \hat{h}''}{\Delta}\right) \right), \nn \\[2mm]
\hat{f}_3 & = \frac{c}{4 L^2} \Bigg[d \left( \frac{ \hat{h} \hat{h}''}{ \Delta} \left( 2 \hat{h}- (x_2-l) \hat{h}' \right) \right) -\left(\hat{h}'' -(x_2-l) \hat{h}'''  \right) x_1 dx_1  - (x_2-l) \hat{h}'' dx_2  \Bigg]  \wedge J_2 , \nn \\[2mm]
\hat{f}_5 & = -\frac{c}{m} \Bigg[ x_1 \left((x_2-l) \left(2 \hat{h}''-(x_2-l) \hat{h}''' \right)+2 \hat{h}' \right) dx_1 +(2 \hat{h}+ (x_2-l)^2 \hat{h}'') dx_2  \nn \\[2mm] 
& + d\left(\frac{\hat{h}}{\Delta} \left((x_2-l) \hat{h}'' \left( 4 \hat{h} - (x_2-l) \hat{h}' \right) - 2 \hat{h} \hat{h}' \right) \right) \Bigg] \wedge J_2 \wedge J_2. \label{andrea-class2RR}
\end{align}
The electric ones, that will be needed for the field theory analysis, are collected in appendix \ref{electricfluxsol}.

One can check that these are the solutions summarised in section \ref{NATD-CP3},  upon
\begin{equation}
x_1 \to \frac{\rho}{\pi},\qquad x_2 \to r, \qquad L^2 \to \frac{\pi}{2}, \qquad c \to 4 \pi^3. \label{eq:NATDmap2}
\end{equation}

Similarly, a second class of solutions can be obtained by choosing $h=x_2 \hat{h}(x_1)$. The main features of these solutions are presented in appendix \ref{sec:strip2}.

\section{$\text{AdS}_2\times \mathbb{CP}^3\times \text{S}^1\times I$. Field theory perspective}
\label{ATD-CP3}

In this section we discuss the field theory interpretation of the first class of solutions discussed in subsection \ref{annulus} for the Riemann surface an annulus. The field theory dual of the second class of solutions was discussed in \cite{LMP}.

The present class of solutions was constructed in \cite{Conti:2023rul} acting with Abelian T-duality on the family of $\text{AdS}_3\times \mathbb{CP}^3\times I$ solutions to massive Type IIA supergravity found in \cite{Macpherson:2023cbl}. The T-duality was operated with respect to the U(1) acting on the $\text{AdS}_3$ subspace in its parametrisation as a Hopf-fibre over $\text{AdS}_2$, given by \eqref{param}. As emphasized before this is the solution we presented in \eqref{andrea-class1} upon identifying
\begin{equation}
x_1 \to r, \qquad x_2 \to \frac{1}{\pi} y, \qquad L^2 \to \frac{\pi}{2}, \qquad c \to 4 \pi^2, \qquad \hat{h}\to h.
\end{equation}
The explicit form of the solution reads:
\begin{eqnarray}
&& \frac{2}{\pi}ds^2 =\frac{h}{\sqrt{2hh''-(h')^2}}ds^2(\text{AdS}_2)+\frac{8}{h''}\sqrt{2hh''-(h')^2}ds^2(\mathbb{CP}^3)+\frac{\sqrt{2hh''-(h')^2}}{h}\Bigl(dr^2+\frac{1}{\pi^2}dy^2\Bigr)\nonumber \\[2mm]
&& e^{\Phi} =\frac{\sqrt{2(2hh''-(h')^2)}}{\sqrt{h(h'')^3}},\qquad
B=4\pi\Bigl(-(r-l)+\frac{h'}{h''}\Bigr) J_2 +\frac{y}{2}\text{vol}(\text{AdS}_2), \nonumber\\ 
&& \hat{f}_1 =-\frac{h'''}{2\pi}dy, \qquad \hat{F}_3=2(h''-(r-l)h''')\Bigr)dy\wedge J_2 \label{AdS2S1I} \\[2mm] 
&&\hat{f}_5 =-4\pi (2h'+(r-l)(-2h''+(r-l)h'''))dy\wedge J_2\wedge J_2\nonumber \\[2mm]
&&\hat{f}_7 = \frac{16\pi^2}{3}(6h-(r-l)(6h'+(r-l)((r-l)h'''-3h'')))dy\wedge J_2\wedge J_2\wedge J_2 \nonumber
\end{eqnarray}
$h$ is a function of $r$, satisfying the Bianchi identity
\begin{equation}\label{Bianchi}
h''''=0.
\end{equation}
This makes it a cubic function of $r$.
$r$ is the non-compact direction that parametrises the interval in the $\text{AdS}_3\times \mathbb{CP}^3\times I$ solutions, and $y$ is the T-duality direction, normalised such that $y\in [0,2\pi]$.  $\hat{f}_p$ are the magnetic Page fluxes.  $l$ is an integer required to enforce that $B$ lies in the fundamental region when integrated along the $\mathbb{CP}^1$ subspace, which we recall is associated to the creation of NS5-branes,
\begin{equation} \label{fundamental}
\frac{1}{(2\pi)^2}|\int_{\text{2-cycle}}B|\in [0,1].
\end{equation}
This ensures that gauge equivalent configurations are not over counted in the path integral. 
This forces the $r$-direction to be divided in $[l,l+1]$ intervals, with NS5-branes being created in these positions. The previous expressions for $B$ and the Page fluxes are valid for $r\in [l,l+1]$. 

To have a well defined dual theory we must ensure that the interval spanned by $r$ is bounded. As explained in more detail in \cite{Macpherson:2023cbl}, viewed as solutions in massive IIA the interval can terminate from above or below in terms of one of the following physical and fully localised singular behaviours
\beq
\text{KK-monopole},~~~~\text{O2},~~~~ \text{D8/O8}\label{eq:boundaries}
\eeq 
where the \text{KK-monopole} is an object that emerges by dimensional reduction of an 8 dimensional monopole geometry with $\mathbb{C}^4/\mathbb{Z}_k$ singularity. Realising this amounts to imposing one of 
\begin{subequations}
\begin{align}
\text{D8/O8}&:~~~~h=c_1+c_2(r-r_0)^3,\label{eq:hprofile1}\\[2mm]
\text{KK-monopole}&:~~~~h=(c_1+c_2 (r-r_0))(r-r_0)^2,\label{eq:hprofile2}\\[2mm]
\text{O2}&:~~~~h=c_1+c_2  (r-r_0)+ \frac{c_2^2}{4 c_1} (r-r_0)^2+c_3 (r-r_0)^3,\label{eq:hprofile3}
\end{align}
\end{subequations}
where $r_0$ is the upper/lower bound of the final/initial $r\in [l,l+1]$ cell and $c_i$ are non zero constants.
 There are several options for constructing solutions with bounded intervals in terms of these behaviours. The simplest is probably to start the space at $r=0$ with one of the above behaviours, place D8 branes at the points $r=l$ along the interval until one reaches the arbitrary point $r=P$ where one then glues the profile for $h$ up to this point onto its  onto its mirror image at $r=P+1$ as explained in section 2.2 of \cite{Lozano:2024idt} - the space then begins and ends with the same singularity. T-dualising these solutions to type IIB does not change the profile for $h$ but the boundary behaviours become
\beq
\text{KK-monopole},~~~~\text{O1},~~~~ \text{D7/O7},\label{eq:boundariesIIB}
\eeq 
where the new KK-monopole in IIB is the result of T-dualising the IIA version on one of its world-volume directions.

The existence of this class of solutions motivated the general classification pursued in this work. Being T-dual to the $\text{AdS}_3\times \mathbb{CP}^3\times I$ solutions in \cite{Macpherson:2023cbl} it is however guaranteed that they are dual to the same CFT, this time realised as a superconformal quantum mechanics. The 2d realisation was studied in  \cite{Lozano:2024idt}. We will summarise its main ingredients in this section, closely following that paper, to which the reader is referred for more details. This discussion will serve as the starting point for our constructions in section \ref{NATD-CP3}, where the dual quantum mechanics of more general $\text{AdS}_2\times \mathbb{CP}^3$ solutions will be studied.

As for the $\text{AdS}_3\times \mathbb{CP}^3\times I$ solutions, the quantum mechanics dual to the solutions \eqref{AdS2S1I}  is better understood  from the brane set-up that is obtained after (Abelian) T-dualising over $\psi$, the Hopf fibre of the $\text{S}^3$ contained in the $\mathbb{CP}^3$, in its parametrisation as a foliation over a $\mathbb{T}^{1,1}$. This gives rise to a Type IIA brane realisation that we have depicted in Table \ref{table1}. This intersection can also be obtained by T-dualising along the common spatial direction the brane set-up associated to the $\text{AdS}_3\times \mathbb{CP}^3\times I$ solutions in their Type IIB realisation, that we have depicted in Table \ref{table2}. 
\begin{table}[h]
\renewcommand{\arraystretch}{1}
\begin{center}
\scalebox{1}[1]{
\begin{tabular}{c| c cc  c c  c  c c c c}
 branes & $x^0$ & $y$ & $r$ & $x^3$ & $x^4$ & $x^5$ & $\psi$ & $x^7$ & $x^8$ & $x^9$ \\
\hline \hline
$\mrm{D}2$ & $\times$ & $-$ & $\times$ & $-$ & $-$ & $-$ & $\times$ & $-$ & $-$ & $-$ \\
$\mrm{NS}5'$ & $\times$ & $\times$ & $\times$ & $\times$ & $\times$ & $\times$ & $-$ & $-$ & $-$ & $-$ \\ 
$(\mrm{NS}5',\mrm{D}4)$ & $\times$ & $(\times,-)$ & $\times$ & $\cos{\phi}$  & $\cos{\phi}$  & $\cos{\phi}$  & $-$ & $\sin{\phi}$ & $\sin{\phi}$ & $\sin{\phi}$ \\
$\mrm{D}6$ & $\times$ & $-$ & $-$ & $\times$ & $\times$ & $\times$ & $-$ & $\times$ & $\times$ & $\times$ \\
$\mrm{NS}5$ & $\times$ & $\times$ & $-$ & $-$ & $-$ & $-$ & $\times$ & $\times$ & $\times$ & $\times$ \\
$\mrm{F}1$ & $\times$ & $\times$ & $-$ & $-$ & $-$ & $-$ & $-$ & $-$ & $-$ & $-$ \\
\end{tabular}
}
\caption{Brane intersection associated to the $\text{AdS}_2\times \mathbb{CP}^3\times \text{S}^1\times I$ solutions \eqref{AdS2S1I}. The intersection describes a brane box \cite{Hanany:2018hlz} with D2 colour branes extended along the $r$ and $\psi$ directions. The (NS5$'$,D4) bound states are rotated the same angle on the $[3,7]$, $[4,8]$ and $[5,9]$ directions with respect to the NS5$'$-branes, giving rise to $\mathcal{N}=3$ supersymmetry in 1d. Fractional D2-branes stretched between $\psi=\pi$ and $\psi=2\pi$ add up to the number of D2-branes in this interval.} \label{table1}
\end{center}
\end{table}
\begin{table}[h]
\renewcommand{\arraystretch}{1}
\begin{center}
\scalebox{1}[1]{
\begin{tabular}{c| c cc  c c  c  c c c c}
 branes & $x^0$ & $x^1$ & $r$ & $x^3$ & $x^4$ & $x^5$ & $\psi$ & $x^7$ & $x^8$ & $x^9$ \\
\hline \hline
$\mrm{D}3$ & $\times$ & $\times$ & $\times$ & $-$ & $-$ & $-$ & $\times$ & $-$ & $-$ & $-$ \\
$\mrm{NS}5'$ & $\times$ & $\times$ & $\times$ & $\times$ & $\times$ & $\times$ & $-$ & $-$ & $-$ & $-$ \\
$(1,k) 5'$ & $\times$ & $\times$ & $\times$ & $\cos{\theta}$ & $\cos{\theta}$ & $\cos{\theta}$ & $-$ & $\sin{\theta}$ & $\sin{\theta}$ & $\sin{\theta}$ \\
$\mrm{D}7$ & $\times$ & $\times$ & $-$ & $\times$ & $\times$ & $\times$ & $-$ & $\times$ & $\times$ & $\times$ \\
$\mrm{NS}5$ & $\times$ & $\times$ & $-$ & $-$ & $-$ & $-$ & $\times$ & $\times$ & $\times$ & $\times$ \\
\end{tabular}
}
\caption{Brane intersection associated to the $\text{AdS}_3\times \mathbb{CP}^3\times I$ solutions in \cite{Macpherson:2023cbl} in its Type IIB realisation \cite{Lozano:2024idt}. The intersection describes a brane box with D3 colour branes extended along the $r$ and $\psi$ directions. The $(1,k)$5$'$-branes are rotated the same angle on the $[3,7]$, $[4,8]$ and $[5,9]$ directions with respect to the NS5$'$-branes, giving rise to $\mathcal{N}=(0,3)$ supersymmetry in 2d. Fractional D3-branes extended between $\psi=\pi$ and $\psi=2\pi$ add up to the number of D3-branes in this interval.} \label{table2}
\end{center}
\end{table}

In the brane set-up described by Table \ref{table1} supersymmetry is first reduced to $\mathcal{N}=4$, due to the T-duality transformation onto Type IIA along the $\psi$ direction, and then to $\mathcal{N}=3$, due to the rotations between the branes.
Still, it is expected that the $\mathcal{N}=3$ theory has the same field content as that of an $\mathcal{N}=4$ quantum mechanics, except for the deformations introduced by the rotations of the branes. In particular, the $\mathcal{N}=4$ theory has SO(4) R-symmetry, that should reduce to SO(3)$_R = \text{diag}(\text{SU}(2)\times
\text{SU}(2))$ for the $\mathcal{N}=3$ theory.
Further, it is expected that, as in 3d, the 1d theory flows to a $\mathcal{N}=4$ CFT in the IR, in its Type IIA realisation, and to a $\mathcal{N}=6$ theory in Type IIB, once the (Abelian) T-duality along $\psi$ is undone, consistently with the number of supersymmetries preserved by the solutions. 

\begin{figure}[h]
\centering
\includegraphics[scale=0.6]{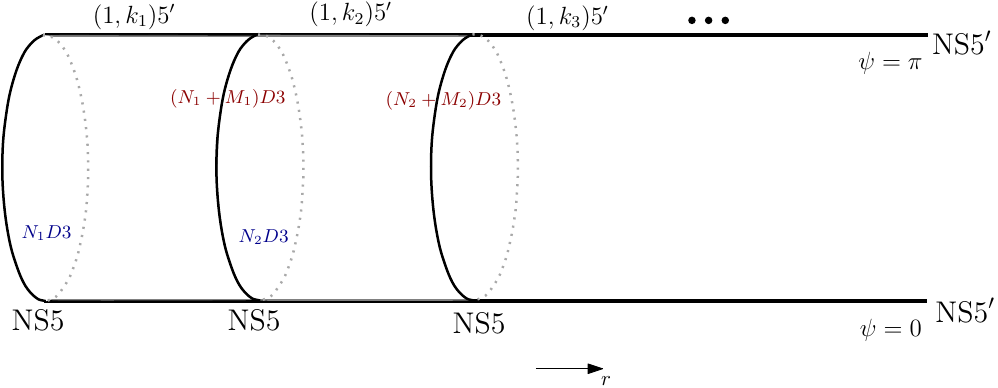}
\caption{Brane configuration associated to the $\text{AdS}_3\times\mathbb{CP}^3\times I$ solutions. $N_l$ D3-branes stretch between $\psi=0$ and $\pi$ and $N_l+M_l$ between $\psi=\pi$ and $2\pi$. The details of the construction can be found in \cite{Lozano:2024idt}.}\label{branediagram}
\end{figure}

Given that $\mathcal{N}=4$ quantum mechanics can be built out of 2d (0,4) multiplets\footnote{$\mathcal{N}=4$ multiplets originate from 2d $(0,4)$ multiplets upon dimensional reduction (see for instance \cite{Okazaki:2015pfa}).}, the quiver that describes the quantum mechanics dual to the $\text{AdS}_2\times \mathbb{CP}^3\times \text{S}^1\times I$ solutions is in fact identical to the one associated to the $\text{AdS}_3\times \mathbb{CP}^3\times I$ solutions. 
Recalling their brane picture, depicted in Figure \ref{branediagram},
we have that for the $\text{AdS}_2$ solutions the D3-branes become D2-branes, that play the role of new colour branes, and that the bound states at $\psi=\pi$ are now built out of an NS5$'$-brane and $k_l$ D4-branes, and are rotated with respect to the NS5$'$-brane at $\psi=0,2\pi$ an angle $\theta_l$ that varies along the $r$-direction\footnote{Even if this is not shown explicitly in the figure for the sake of clarity.}. Besides this, the role played by the D7-branes is now played by D6-branes. In more detail, in the brane diagram $N_l$ D2-branes are bounded between NS5-branes along the $r$ direction, and between the NS5$'$ brane located at $\psi=0$ and the $(\text{NS5}', k_l \,\text{D4})$ branes located at $\psi=\pi$ along the $\psi$-direction. This number becomes $N_l+M_l$ between the $(\text{NS5}', k_l \,\text{D4})$  branes and the NS5$'$-brane at $\psi=2\pi$, due to extra fractional D3-branes created between these branes.

\begin{figure}[h]
\centering
\includegraphics[scale=0.65]{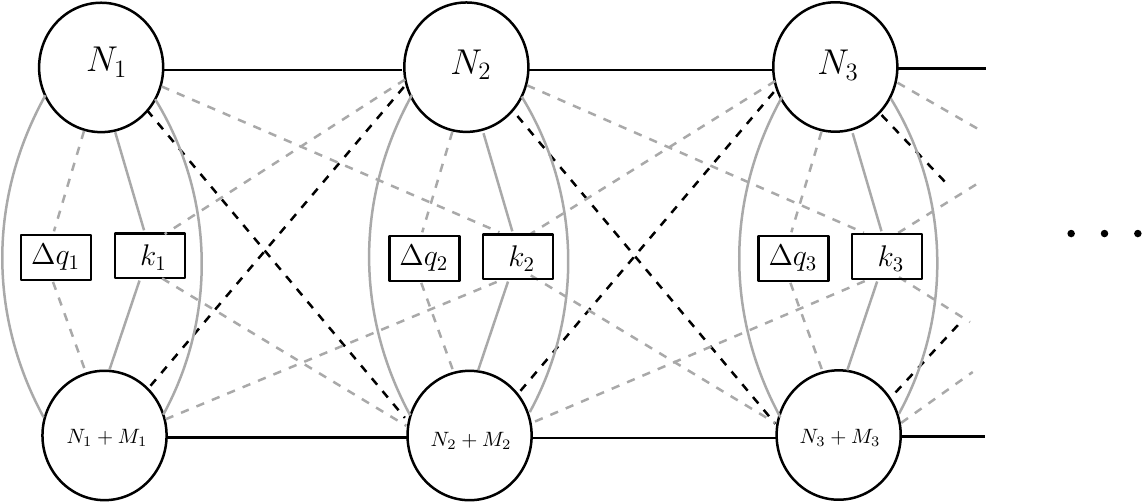}
\caption{Quiver diagram associated to the $\text{AdS}_2\times \mathbb{CP}^3\times \text{S}^1\times I$ solutions. Circles denote $\mathcal{N}=4$ vector multiplets, black lines $\mathcal{N}=4$ twisted hypermultiplets, grey lines $\mathcal{N}=4$ hypermultiplets, grey dashed lines $\mathcal{N}=2$ Fermi multiplets and black dashed lines $\mathcal{N}=4$ Fermi multiplets.}\label{quiverATD}
\end{figure}

The associated quiver is thus identical to the one constructed in \cite{Lozano:2024idt}, depicted in Figure \ref{quiverATD}. We have chosen to describe it in terms of 1d $\mathcal{N}=4$ multiplets even if these simply arise from 2d $(0,4)$ ones upon dimensional reduction.
Since $\psi$ is a compact direction, there are two $\mathcal{N}=4$ hypermultiplets connecting the $N_l$ and $N_l+M_l$ gauge groups at each $r$-interval, that originate from the open strings that connect the $N_l$ D2 and $N_l+M_l$ D2 branes across either one of the NS5$'$ branes. Similarly, there are two $\mathcal{N}=2$ Fermi multiplets connecting $N_l$ D2-branes with $N_{l'}+M_{l'}$ D2 branes in adjacent $r$-intervals, depending on the NS5$'$-brane crossed by the open strings. In each $r$-interval the two $\mathcal{N}=4$ bifundamental hypermultiplets connecting the gauge nodes with ranks $N_l$ and $N_l+M_l$  combine onto a $\mathcal{N}=6$ bifundamental hypermultiplet. On top of this we have the $\mathcal{N}=4$ fundamental hypermultiplets and $\mathcal{N}=2$ fundamental Fermi multiplets associated to the open strings that connect the D2-branes with the D4-branes and the D6-branes. In all, the quiver field theory consists on a sequence of U$(N_l)$ and U$(N_l+M_l)$ gauge groups with the field content of $\mathcal{N}=4$ vector multiplets connected to each other by $(0,6)$ bifundamental hypermultiplets, with extra $k_l$ fundamental or antifundamental $\mathcal{N}=4$ hypermultiplets\footnote{Here we have taken into account that due to the different relative positions of the NS5$'$-branes at $\psi=0,2\pi$ compared to the $(\text{NS5}', k_l \,\text{D4})$  branes  at $\psi=\pi$, the flavour groups contribute with fundamentals to the $N_l$ gauge nodes and with antifundamentals to the $N_l+M_l$ gauge nodes.}.  As stressed out in \cite{Lozano:2024idt} in the 2d case, this quiver describes a $\mathcal{N}=4$ quantum mechanics to which the effect of the rotation between the branes must be added, that renders some of the multiplets massive. This has an effect on the computation of the gauge anomaly of the 2d theory. In the 1d case there is no gauge anomaly, but the same calculation holds. The reader is referred to \cite{Lozano:2024idt}  for more details.

In the next section we extend these constructions to more general $\text{AdS}_2\times \mathbb{CP}^3$ solutions where the Riemann surface is a strip, in particular to the non-Abelian T-dual solutions constructed in \cite{Conti:2023rul}.

\section{$\text{AdS}_2\times \mathbb{CP}^3\times I \times I'$. Field theory perspective}\label{NATD-CP3}

In \cite{Conti:2023rul} a further class of solutions in Type IIB supergravity realised as $\text{AdS}_2\times \mathbb{CP}^3$ foliations over a 2d Riemann surface with the topology of a strip was constructed. The solutions were again obtained from the $\text{AdS}_3\times \mathbb{CP}^3\times I$ solutions to massive Type IIA supergravity found in \cite{Macpherson:2023cbl}, this time acting with non-Abelian T-duality with respect to one of the freely acting SL(2,$\mathbb{R}$)'s on the $\text{AdS}_3$ subspace. As emphasised before this is the solution we presented in \eqref{andrea-class2}, \eqref{andrea-class2RR} upon identifying
\begin{equation}
x_1 \to \frac{\rho}{\pi},\qquad x_2 \to r, \qquad L^2 \to \frac{\pi}{2}, \qquad c \to 4 \pi^3, \qquad \hat{h}\to h.
\end{equation}
The explicit form of the solutions reads:
\begin{eqnarray} \label{Page1}
&&\frac{2}{\pi}ds^2=\frac{ h }{\sqrt{2hh''-(h')^2}}\frac{\rho^2}{\Xi}ds^2(\text{AdS}_2)+\frac{8}{h''}\sqrt{2hh''-(h')^2}ds^2(\mathbb{CP}^3) \nonumber \\[2mm]
&& \hspace{1cm} + \frac{ \sqrt{2hh''-(h')^2}}{h} \Bigl(dr^2+\frac{1}{\pi^2}d\rho^2\Bigr), \qquad 
e^{\Phi}=\frac{1}{\sqrt{\Xi}}\frac{\sqrt{2(2hh''-(h')^2)}}{\sqrt{h(h'')^3}},\nonumber \\[2mm]
&&B=4\pi\Bigl(-(r-l)+\frac{h'}{h''}\Bigr) J_2 +\Bigl(\frac{\rho^3}{2\Xi}-n\pi\Bigr)\text{vol}(\text{AdS}_2),~~~~\Xi=\rho^2-\frac{\pi^2h^2}{2hh''-(h')^2} \nonumber \\[2mm]
&&\hat{F}_1=\frac{h'''}{2\pi}\rho d\rho+\hat{F}_1^r, \nonumber \\[2mm]
&&\hat{F}_3=-2(h''-(r-l)h''')\rho \, d\rho \wedge J_2+n\pi \,\text{vol}(\text{AdS}_2) \wedge \hat{F}_1^r+\hat{F}_3^{r,\mathbb{CP}^1}, \nonumber\\[2mm]
&&\hat{F}_5=4\pi  (2h'+(r-l)(-2h''+(r-l)h'''))\rho d\rho\wedge J_2\wedge J_2+n\pi\,\text{vol}(\text{AdS}_2) \wedge \hat{F}_3^{r,\mathbb{CP}^1}+\hat{F}_5^{r,\mathbb{CP}^2}, \nonumber\\[2mm]  
&&\hat{F}_7=\frac{16\pi^2}{3} (6h-(r-l)(6h'+(r-l)((r-l)h'''-3h'')))\rho d\rho\wedge J_2\wedge J_2\wedge J_2, \nonumber\\[2mm]  
&&\hspace{1cm}+n\pi\,\text{vol}(\text{AdS}_2) \wedge \hat{F}_5^{r,\mathbb{CP}^2} +\hat{F}_7^{r,\mathbb{CP}^3}, \nonumber\\[2mm]
&&\hat{F}_9=n\pi\,\text{vol}(\text{AdS}_2) \wedge \hat{F}_7^{r,\mathbb{CP}^3}. 
\end{eqnarray}
In these expressions $\rho$ is the non-compact variable that arises from the non-Abelian T-duality transformation, that maps the SL(2,$\mathbb{R}$) gauge group onto its Lie algebra, that we have parametrised in $(\rho,\text{AdS}_2)$ coordinates to manifestly realise the residual SL(2,$\mathbb{R}$) global symmetries \cite{Ramirez:2021tkd}.  In the expressions for the Page fluxes we have just given the relevant components for the analysis of the dual field theory. The expressions for $\hat{F}_1^r$, $\hat{F}_3^{r,\mathbb{CP}^1}$, $\hat{F}_5^{r,\mathbb{CP}^2}$ and $\hat{F}_7^{r,\mathbb{CP}^3}$, where the superscripts denote the respective components, and the extra terms required by Hodge duality can be extracted from \eqref{andrea-class2RR} and \eqref{eq:genelpages} after identifying \eqref{eq:NATDmap2} and \eqref{eq:NATDmap1}. $l$ and $n$ are integers required to enforce that $B$ lies in the fundamental region when  integrated over the $\mathbb{CP}^1$ or the $\text{AdS}_2$ subspaces. We recall this is associated to the creation of  NS5-branes and F1-strings. This forces the $r$-direction to be divided in $[l,l+1]$ intervals, as for the Abelian T-dual solution, and the $\rho$-direction to be divided in $[2n\pi,2(n+1)\pi]$ intervals, with F1-strings being created stretched between $\rho_n=2n\pi$ and $\rho_{n+1}=2(n+1)\pi$. The expressions in \eqref{Page1} for $B$ and the Page fluxes hold for $r\in [l,l+1]$ and $\rho\in [2n\pi,2(n+1)\pi]$. 

If we choose to non-Abelian T-dualise a solution for which the interval spanned by $r$ was bounded in type IIA, as described around \eqref{eq:boundaries}, then it remains so in IIB and one expects one of the behaviours \eqref{eq:boundariesIIB} bounding the interval at each end. However the T-dual coordinate $\rho$ is not compact like $y$ was in \eqref{AdS2S1I}, we can determine its domain by considering the function $\Xi$ which appears in the metric, dilaton and NS 2-form: $\Xi$ becomes negative for $\rho^2<\frac{\pi^2h^2}{2hh''-(h')^2}$ so the $\rho$ interval is bounded from below at $\rho=\sqrt{\frac{\pi^2h^2}{2hh''-(h')^2}}$. As $\Xi$ neither tends to zero nor blows up as we move towards $\rho\to \infty$  the $\rho$ coordinate is semi infinite, as is typical for non-Abelian T-dual solutions. As one approaches $\rho=\infty$ the metric becomes locally that of \eqref{AdS2S1I} but $e^{\Phi}$ is vanishing and so exhibits singular behaviour that we are unable to identify as physical. That $\rho$ spans a semi-finite interval is our first indication that the non-Abelian T-dual solutions need to be ``completed'' in the fashion of \cite{Lozano:2016kum,Lozano:2016wrs}. The lower bound of the $\rho$ interval depends on $r$ as $\rho=\frac{\pi^2h^2}{2hh''-(h')^2}$, so the behaviour at this locus depends where we are along the $r$ interval. At generic points along $r$, where is a $\frac{\pi^2h^2}{2hh''-(h')^2}$ finite constant, one has an OF1 plane singularity at the lower bound of $\rho$, which is the S-dual of a O1 plane. At the end points of the  $r$ interval however we have singular behaviour, and in particular tuning $h$ in the upper/lower cell as either \eqref{eq:hprofile1} or \eqref{eq:hprofile3} means that $\frac{\pi^2h^2}{2hh''-(h')^2}$ blows up at this loci contracting the interval spanned by  $\rho$ to a point. As this does not fit well with a division of the $\rho$ interval into unit length cells we discard this possibility and take the $r$ interval to terminate between monopoles by taking $h$ to have the behaviour of \eqref{eq:hprofile2}. At generic points along the $\rho$ interval and as $\rho\to \infty$ the solution does indeed realise a monopole geometry at the boundaries of $r$, but if we are simultaneously at the lower bound of $\rho$ and an end point of $r$ things are a bit more subtle. We have that
\beq
\frac{\pi^2h^2}{2hh''-(h')^2}\to \frac{c_1\pi^2}{4c_2}(r-r_0)
\eeq
as we approach the monopole, so both $(r-r_0)$ and $\rho$ tend to zero as we approach such a point. We are also unable to make physical sense of this singularity, a second sign that the non-Abelian T-dual requires a completion such that the internal space is bounded between physical behaviours. Our  general class of ${\cal N}=6$ AdS$_2$ solutions should provide such a completion, similarly to how the ${\cal N}=2$ (${\cal N}=4$) AdS$_5$ (AdS$_4$) class of \cite{Gaiotto:2009gz} (\cite{DHoker:2007zhm,DHoker:2007hhe}) did for the non-Abelian T-dual solutions of \cite{Lozano:2016kum} (\cite{Lozano:2016wrs}). The general idea is that the non-abelian T-dual is describing a linear quiver of infinite length, but by embedding it into a more general class of solutions one can make this quiver finite and its dual geometry compact. This generalised solution should share many general  features of the non-Abelian T-dual solution, such as yielding the same Page charges at generic points in the space, but be well defined globally. However the success of previous completions depended on not only the existence of a more general class but also a detailed understanding of the well defined solutions it contained. We aim to gain such an understanding of ${\cal N}=6$ AdS$_2$ solutions in \cite{CLM}, but we currently lack this. Thus for the rest of this section, guided by the Page charges, we will focus on the superconformal quantum mechanics that the non-Abelian T-dual suggests should exist, but will stop short of constructing the actual well defined geometry it is dual to.
 
As for the solutions described in the previous section, the existence of the solutions \eqref{Page1} suggested that a general class of Type IIB $\mathcal{N}=6$ supersymmetric $\text{AdS}_2\times \mathbb{CP}^3$ geometries, this time foliated over a 2d Riemann surface with the topology of a strip, should exist. We have constructed this general class in this paper. However, in contrast with the solutions described in the previous section, the solutions \eqref{Page1}, being generated through non-Abelian T-duality,  are not expected to describe the same CFT as the original  $\text{AdS}_3\times \mathbb{CP}^3\times I$ solutions. Therefore, they represent a richer class  from the field theory perspective. Indeed, existing examples in the literature in different dimensions suggest that the dual field theory changes, both under SU(2) \cite{Itsios:2013wd}-\cite{Lozano:2019ywa} and SL(2,$\mathbb{R}$) \cite{Lozano:2021rmk,Ramirez:2021tkd} non-Abelian T-dualities. We will see that our current geometries provide one further example of this feature.

\begin{figure}[h]
\centering
\includegraphics[scale=0.65]{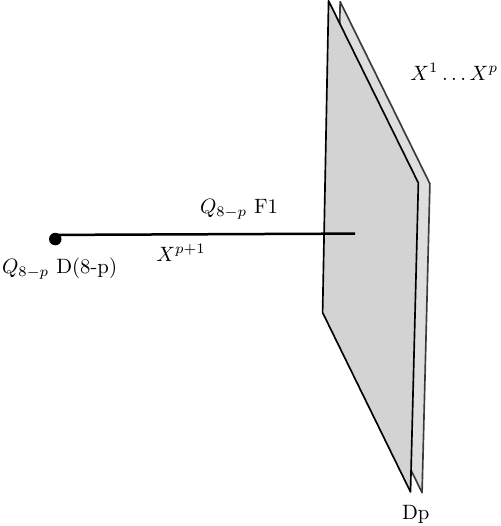}
\caption{F1-strings stretched between Dp and D(8-p) branes.}\label{Wilson-lines}
\end{figure}

We start discussing the associated brane set-up. The brane intersection suggested from the NS-NS and RR fluxes in \eqref{Page1} consists on the already identified brane set-up associated to the Abelian T-dual solutions, depicted in Table \ref{table1}, with possibly extra D6$'$, D4$'$ and D2$'$ branes orthogonal to the D2 and F1, D4 and F1 and D6 and F1 branes, respectively. In the Type IIB description the D$p$-branes are charged with respect to the $\hat{F}_{8-p}^{\rho,\mathbb{CP}^{(7-p)/2}}$ components of the Page fluxes in \eqref{Page1}, while the D$(8-p)'$-branes would be charged with respect to the $\hat{F}_p^{r,\mathbb{CP}^{(p-1)/2}}$ components. The coupling
\begin{equation}\label{Wilson-couplings}
S_{Dp}=-T_{p}\int_{(t,I_r,\mathbb{CP}^{(p-1)/2})}\hat{F}_{p}\wedge A_t=-T_{F1}Q'_{8-p}\int A_t,
\end{equation}
in the worldvolume effective action of the D$p$-branes shows that indeed the latter fluxes induce electric charge on them. 
Figure \ref{Wilson-lines} illustrates one such D$p$-F1-D$(8-p)$ Wilson line configuration. These configurations have been studied in \cite{Tong:2014cha,Assel:2018rcw,Lozano:2020sae,Lozano:2021rmk,Ramirez:2021tkd,Lozano:2022vsv}, and they have been shown to arise, in particular, when a non-Abelian T-duality transformation is performed with respect to a freely acting SL(2,$\mathbb{R}$) on AdS$_3$ (see \cite{Ramirez:2021tkd} for a detailed account on this).
In this setting  the effect of the non-Abelian T-duality transformation is to create Wilson lines on the subset of branes already generated upon dualising with respect to the U(1) $\subset$ SL(2,$\mathbb{R}$) subgroup \cite{Lozano:2021rmk,Ramirez:2021tkd}. The F1-strings that emerge after the non-Abelian T-duality extend along the $\rho$-direction, which is non-compact, and they need to terminate in new branes, orthogonal to both the ``Abelian T-dual branes'' and the F1's. In the detailed example discussed in 
\cite{Ramirez:2021tkd}, where the non-Abelian T-duality was performed on the $\text{AdS}_3$ subspace of the $\text{AdS}_3\times \text{S}^3\times\text{CY}_2$ Type IIB background, the extra branes were D8 and D4$'$ branes, orthogonal, respectively, to the D0 and D4 branes that arise already when dualising with respect to the U(1) fibre of the $\text{AdS}_3$ space. On top of these branes, coming from the dualisation of the D1-D5 system, there are extra F1-strings that in the non-Abelian T-dual case extend along the $\rho$ non-compact direction. In the current example, and going back to the Type IIA description, the D6 and D2 branes must already play the role of orthogonal branes with respect to each other, since this is the only possibility consistent with the preserved supersymmetries. Instead, for the D4-branes there are extra D4$'$-branes orthogonal to them and to the F1-strings. 
This is compatible with our previous discussion of the Page fluxes because one expects the $\rho$ and $r$ directions and the radius of $\text{AdS}_2$ to mix with each other in the underlying brane intersection. It would be very interesting to construct the brane intersection that underlies the solutions to further clarify this issue. This however would imply not only identifying the brane intersection underlying the original $\text{AdS}_3\times \mathbb{CP}^3\times I$ solutions, previous to the non-Abelian T-duality transformation, but also to understand how it transforms under non-Abelian T-duality. This remains however an open issue, even in simpler brane configurations (see \cite{Terrisse:2018hhf}). Still, based on the preservation of supersymmetries and the information provided by the $\text{AdS}_2$ near-horizon geometry we can propose the brane set-up depicted in Table \ref{table3} as underlying the class of $\text{AdS}_2\times \mathbb{CP}^3\times I\times I'$ solutions \eqref{Page1} (again, in its Type IIA realisation). Note that we are still using $\rho$ and $r$ variables  to try to reflect as much as possible the information provided by the near horizon limit. 
\begin{table}[h]
\renewcommand{\arraystretch}{1}
\begin{center}
\scalebox{1}[1]{
\begin{tabular}{c| c cc  c c  c  c c c c}
 branes & $x^0$ & $\rho$ & $r$ & $x^3$ & $x^4$ & $x^5$ & $\psi$ & $x^7$ & $x^8$ & $x^9$ \\
\hline \hline
$\mrm{D}2$ & $\times$ & $-$ & $\times$ & $-$ & $-$ & $-$ & $\times$ & $-$ & $-$ & $-$ \\
$\mrm{NS}5'$ & $\times$ & $\times$ & $\times$ & $\times$ & $\times$ & $\times$ & $-$ & $-$ & $-$ & $-$ \\ 
$(\mrm{NS}5',\mrm{D}4)$ & $\times$ & $(\times,-)$ & $\times$ & $\cos{\theta}$  & $\cos{\theta}$  & $\cos{\theta}$  & $-$ & $\sin{\theta}$ & $\sin{\theta}$ & $\sin{\theta}$ \\
$\mrm{D}6$ & $\times$ & $-$ & $-$ & $\times$ & $\times$ & $\times$ & $-$ & $\times$ & $\times$ & $\times$ \\
$\mrm{NS}5$ & $\times$ & $\times$ & $-$ & $-$ & $-$ & $-$ & $\times$ & $\times$ & $\times$ & $\times$ \\
$\mrm{F}1$ & $\times$ & $\times$ & $-$ & $-$ & $-$ & $-$ & $-$ & $-$ & $-$ & $-$ \\
\hline
$\mrm{D}4'$ & $\times$ & $-$ & $-$ & $-\sin{\theta}$ & $-\sin{\theta}$ & $-\sin{\theta}$ & $\times$ & $\cos{\theta}$ & $\cos{\theta}$ & $\cos{\theta}$ \\
\end{tabular}
}
\caption{Brane intersection associated to the solutions \eqref{Page1}. The intersection describes a brane box with D2 colour branes extended along the $r$ and $\psi$ directions with Wilson lines added along the $\rho$-direction. The (NS5$'$,D4) branes are rotated the same angle on the $[3,7]$, $[4,8]$ and $[5,9]$ directions with respect to the NS5$'$-branes, giving rise to $\mathcal{N}=3$ supersymmetry in 1d. Extra fractional branes extend between $\psi=\pi$ and $\psi=2\pi$ that add up to the number of D2-branes in this interval. Wilson lines extend along the $\rho$-direction, stretching between the D2 and the D6 branes and between the D4 and the D4$'$ branes. The latter must be orthogonal to the (NS5$'$,D4) bound states.} \label{table3}
\end{center}
\end{table}

The brane set-up depicted in Table \ref{table3} suggests that the gauge theory associated to the solutions should live in D2-branes stretched between NS5 and NS5$'$ branes positioned along the $r$ and $\psi$ directions, as in the $\text{AdS}_2\times \mathbb{CP}^3\times \text{S}^1\times I$ class of solutions discussed in the previous section. The  components of the Page fluxes along the $\rho$, $(\rho, \mathbb{CP}^1)$, $(\rho,\mathbb{CP}^2)$ and $(\rho,\mathbb{CP}^3)$ directions give rise to the following quantised charges\footnote{In the Type IIA description the D1, D5 and D7 brane charges become, respectively, D2, D4 and D6 brane charges. There are as well extra fractional branes that add up to the D2-brane charges.}
\begin{eqnarray}
Q_1^{(n)}&=&\frac{1}{(2\pi)^6}\int \hat{F}_7^{\rho,\mathbb{CP}^3}=(2n+1)Q_{1}^{(0)} \label{charge1}\\
Q_3^{(n)}&=&\frac{1}{(2\pi)^4}\int \hat{F}_5^{\rho,\mathbb{CP}^2}=(2n+1)Q_3^{(0)}\\
Q_5^{(n)}&=&\frac{1}{(2\pi)^2}\int \hat{F}_3^{\rho,\mathbb{CP}^1}=(2n+1)Q_5^{(0)}\\
Q_7^{(n)}&=&\int \hat{F}_1^\rho=(2n+1)Q_7^{(0)} \label{charge4},
\end{eqnarray}
where $Q_p^{(0)}$ are the charges in the $\rho\in [0,2\pi]$ interval. These charges depend on the specific form of the function $h(r)$. As shown in \cite{Lozano:2024idt} the most general global solutions satisfying the Bianchi identity \eqref{Bianchi} are constructed by glueing local solutions with D7-branes placed at $r=l$. This means that $h$ needs only be piecewise cubic.
As in \cite{Lozano:2024idt} the four parameters on which $h$ depends can be chosen to be $(Q_{1}^{(0)}, Q_{3}^{(0)}, Q_{5}^{(0)}, Q_{7}^{(0)})$, in each $r$-interval, using the explicit expressions of the $\hat{F}_p^{\rho,\mathbb{CP}^{(p-1)/2}}$  Page fluxes. They can be determined from the number of D7-branes in each interval imposing continuity to $(h,h',h'')$. Doing that and introducing a subscript $l$ to explicitly account for the fact that they depend on the $r$-interval, one finds
\begin{equation}\label{hfunction}
\pi h_l(r)=Q_{1,l}^{(0)}-Q_{3,l}^{(0)}(r-l)+\frac12 Q_{5,l}^{(0)}(r-l)^2-\frac16 Q_{7,l}^{(0)}(r-l)^3 \qquad \text{for} \qquad r\in[l,l+1].
\end{equation}
Continuity of $(h,h',h'')$ then imposes that the quantised charges should change across $r$-intervals as
\begin{eqnarray}
&&Q_{1,l}^{(0)}=Q_{1,l-1}^{(0)}-Q_{3,l-1}^{(0)}+\frac12 Q_{5,l-1}^{(0)}-\frac16 Q_{7,l-1}^{(0)} \nonumber \\
&&Q_{3,l}^{(0)}=Q_{3,l-1}^{(0)}-Q_{5,l-1}^{(0)}+\frac12 Q_{7,l-1}^{(0)}\\
&&Q_{5,l}^{(0)}=Q_{5,l-1}^{(0)}-Q_{7,l-1}^{(0)}. \nonumber
\end{eqnarray}
Comparing to the expressions of the quantised charges of the $\text{AdS}_3\times \mathbb{CP}^3\times I$ solutions, prior to the non-Abelian T-duality transformation, one can see that they have changed by a factor of $\pi$ \footnote{This can be seen for instance comparing \eqref{hfunction} with equation (2.30) in  \cite{Lozano:2024idt}.}. This is a common feature under non-Abelian T-duality (see for instance the discussion in \cite{Lozano:2013oma}), contrary to what happens in the Abelian case.
Moreover, the quantised charges depend
on the $\rho$-interval where the branes are located, as shown by equations \eqref{charge1}-\eqref{charge4}.

\begin{figure}[h]
\centering
\includegraphics[scale=0.75]{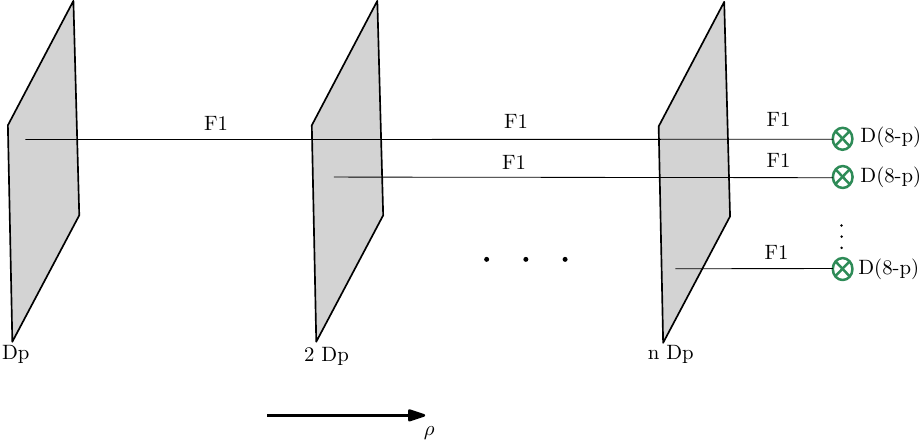}
\caption{F1-strings stretched between the Dp and D(8-p) branes along the $\rho$-direction, in units of $Q'_{8-p}$.}\label{Wilson-lines-total}
\end{figure}

The magnetic charges just discussed are useful to describe the field theory along the $r$-direction. However, on top of these the branes carry as well electric charges, induced by the F1-strings stretched between them and the corresponding orthogonal D$(8-p)$-branes. The latter carry, in turn, magnetic charge with respect to the $\hat{F}_p^{r,\mathbb{CP}^{(p-1)/2}}$ components of the Page fluxes, as implied by the couplings in \eqref{Wilson-couplings}. We have denoted these charges with a prime to distinguish them from the quantised charges computed in \eqref{charge1}-\eqref{charge4}. More specifically, we have for the Page fluxes in \eqref{Page1},
\begin{equation}
\hat{F}_{p+2}^{(\text{AdS}_2,r,\mathbb{CP}^{(p-1)/2})}=n\pi \hat{F}_p^{(r,\mathbb{CP}^{(p-1)/2})}\wedge \text{vol}_{\text{AdS}_2},
\end{equation}
from where
\begin{equation}
Q_{p}^{(e)}=nQ'_{8-p} \qquad \text{for} \qquad \rho\in [2n \pi,2(n+1)\pi].
\end{equation}
Therefore, the picture that arises for the ``electric part'' of the brane set-up is that there are stacks of D$p$-branes located at $\rho_n=2n\pi$ with $n Q'_{8-p}$ F1-strings ending on them that originate from $n$ stacks of D$(8-p)$-branes. This is depicted in Figure \ref{Wilson-lines-total}.
This configuration describes Wilson lines in the $Q'_{8-p}$ completely antisymmetric representation of the gauge groups U$(nQ'_{8-p})$. Therefore it actually describes baryon vertices. 

The global picture that arises is one in which the quiver depicted in Figure \ref{quiverATD} is realised in each $\rho\in [2n\pi,2(n+1)\pi]$ interval, with the number of branes being computed from the quantised charges \eqref{charge1}-\eqref{charge4} in each such interval. Thus, in the non-Abelian T-dual setting the F1-strings induced by the T-duality transformation acquire an interpretation in the field theory together with extra branes (D4$'$ in this case) as describing baryon vertices for the branes that were already present after Abelian T-duality, whose numbers depend now on the $\rho$-interval. This is analogous, although more involved due to the higher complexity of the current backgrounds, to the situation encountered in  \cite{Lozano:2021rmk,Ramirez:2021tkd}, that also analysed backgrounds generated via non-Abelian T-duality with respect to a freely acting SL(2,$\mathbb{R}$) on AdS$_3$. It would be interesting to further study the implications of these realisations, in particular in connection with the description of defects within AdS/CFT, along the lines mentioned in the introduction. 

Finally, we need to recall that in non-trivial manifolds like the $\mathbb{CP}^3$ there are non-trivial effects such as the Freed-Witten anomaly \cite{Freed:1999vc} and higher curvature couplings \cite{Bergman:2009zh} that induce higher dimensional brane charge in the worldvolume of branes, and thus contribute to the final number of these branes in each interval. These effects need to be properly taken into account in order to compute the ranks of the gauge groups in the quiver, as well as the charges associated to flavour branes. A detailed account on these effects is taken in \cite{Lozano:2024idt}, to which the reader is referred for more details. Here we just recall the relation between the quantised charges computed from the RR-fluxes and the actual numbers of branes in each interval, after the Freed-Witten and higher curvature terms have been taken into account \cite{Aharony:2009fc,Bergman:2010xd},
\begin{eqnarray}
&&Q_{1,l}^{(n)}=N_l^{(n)}+\frac{1}{12} k_l^{(n)}\nonumber\\
&&Q_{3,l}^{(n)}=M_l^{(n)}-\frac12 k_l^{(n)}+\frac{1}{12}q_l^{(n)}\nonumber\\
&&Q_{5,l}^{(n)}=k_l^{(n)}\nonumber\\
&&Q_{7,l}^{(n)}=-q_l^{(n)}
\end{eqnarray}

\subsection{Holographic central charge}

The holographic central charge for the general class of solutions can be computed from the usual prescription
\begin{equation}
c_{hol}=\frac{3}{2^5\pi^8}V_{int}=\frac{3}{2^5\pi^8}\int d^8\xi \,e^{-2\Phi}\sqrt{\text{det}(g_{ij})}
\end{equation}
where $g_{ij}$ is the metric on the inner space and $\xi$ coordinates defined over it. This gives
\begin{equation}\label{cholo}
c_{hol}=\frac{3}{32\pi^4 m^2}\int dx_1 dx_2\Bigl(2h\Box h-(\partial_1 h)^2-(\partial_2 h)^2\Bigr).
\end{equation}
This constitutes a non-trivial prediction for the superconformal quantum mechanics dual to the general class of solutions, as this expression should be matched by the field theory result in the holographic limit. The field theory result could be derived from the central extension of the OSp$(6|2)$ superconformal algebra, given by
\begin{equation}
c_R=\frac{k(3k+13)}{k+3}
\end{equation}
where $k$ is the level of the algebra. The level of the algebra has not been computed, to our knowledge, for superconformal quantum mechanics with our preserved supersymmetries. The computation of the central charge from the field theory side remains as an interesting open problem that would be interesting to address. Expression \eqref{cholo} could be used as a lead for these investigations. It is likely that as in the calculation in \cite{Lozano:2024idt}, this expression includes higher derivative contributions, and therefore subleading corrections to the planar free energy.

Regarding the particular subclass of solutions discussed in this section, we find a central charge that differs from the one for the $\text{AdS}_3\times \mathbb{CP}^3\times I$ solutions. Namely, we find
\begin{equation}
c_{1d}= \frac{3 }{2^5 \pi ^4} \int dr d\rho \left(\rho ^2 \left(2 h h''-h'^2\right)-\pi ^2 h^2\right),
\end{equation}
which differs from the expression computed in \cite{Lozano:2024idt} for the $\text{AdS}_3\times \mathbb{CP}^3\times I$ solutions, given by
\begin{equation}
c_{2d}=\frac12 \int dr (2hh''-h'^2)
\end{equation}
This is a common feature under non-Abelian T-duality \cite{Lozano:2013oma,Lozano:2014ata}, which is in agreement with the expectation that it should be dual to a different CFT.

\section{Conclusions}\label{conclusions}

In this paper we have provided a new class of $\text{AdS}_2$ solutions in Type IIB supergravity with $\mathcal{N}=6$ supersymmetry, and work is under way to extend them to $\mathcal{N}=5$ \cite{CLM}. We have shown that the $\mathcal{N}=6$ solutions exist only in Type IIB supergravity, where they are given by $\text{AdS}_2\times \mathbb{CP}^3$ foliations over a two dimensional Riemann surface. The solutions depend on a single function defined on the Riemann surface, which satisfies a Bianchi identity that we have found the general local solution to which can be expressed in terms of two holomorphic functions.

Due to the complexity of the solutions, we have focussed on two subclasses in order to elucidate the main properties of their dual $\mathcal{N}=6$ superconformal quantum mechanics. These are the solutions constructed in \cite{Conti:2023rul}, acting with Abelian and non-Abelian T-duality on the $\text{AdS}_3\times \mathbb{CP}^3\times I$ class presented in \cite{Macpherson:2023cbl}. The existence of these two subclasses of solutions motivated in fact the search for the general class carried out in this paper. We have started with the analysis of the quantum mechanics dual to the Abelian T-dual solution. Given its relation with the $\text{AdS}_3\times \mathbb{CP}^3\times I$ solutions under Abelian T-duality, it is guaranteed that they should be dual to the same CFT. Yet, we have discussed the quantum mechanics in some detail as a way to introduce the ingredients that we will need when we discuss the second subclass of solutions. Even if one could expect that the relation between the CFT duals to solutions related by Abelian T-duality performed along a field theory direction, as in our case, could be cumbersome, this is not the case in our setting given the straightforward relation between 2d $(0,4)$ multiplets and 1d $\mathcal{N}=4$ ones, which are the basic ingredients of the two field theory constructions. 
In turn, the analysis of the field theory dual to the subclass of solutions constructed via non-Abelian T-duality reveals the true nature of the superconformal quantum mechanics dual to the general class of $\text{AdS}_2$ solutions. However, the complexity of the solutions, especially of the RR sector, has prevented us from obtaining a fully general description, but just to give a glimpse on the kind of configurations one could expect. It would be interesting to explore this in more detail.

Likewise, it would be interesting to address in more detail the calculation of the central charge from the field theory perspective. An account on the number of degrees of freedom of the theory is given by the central extension of the superalgebra, in this case OSp$(6|2)$, but to obtain this quantity one needs to know the level of the algebra, which has not been computed, to our knowledge, in the literature. Previous results have shown that for $\mathcal{N}=2$ theories it is possible to use the expression that gives the level of the superalgebra for 2d $(0,2)$ SCFTs in terms of the R-symmetry anomaly, which can be computed away from the conformal fixed point. In the present case, however, the theories preserve $\mathcal{N}=3$ supersymmetries, so the previous prescription cannot be used. This was the same obstacle found in  \cite{Lozano:2024idt} when trying to match the holographic central charge of the $\text{AdS}_3\times \mathbb{CP}^3\times I$ solutions found in \cite{Macpherson:2023cbl} with a field theory result based on $(0,3)$ supersymmetric 2d theories. Clarifying this issue would be an interesting subject for future investigations. 

Finally, it would also be interesting to show whether the holographic central charge can be obtained as a product of electric and magnetic charges associated to the solutions, as found for $\text{AdS}_2$ backgrounds with $\mathcal{N}=4$ supersymmetries in \cite{Lozano:2020txg,Lozano:2020sae,Lozano:2021rmk}. This has been shown to work as well for the $\text{AdS}_3$ solutions in \cite{Lozano:2024idt}, which suggests that the findings in \cite{Lozano:2020txg,Lozano:2020sae,Lozano:2021rmk}  should hold more generally. 

\section*{Acknowledgements}
NTM and YL thank Achilleas Passias for collaboration on related topics. The authors acknowledge support from grants from the Spanish government MCIU-22-PID2021-123021NB-I00 and principality of Asturias SV-PA-21-AYUD/2021/52177. NTM is also supported by the Ram\'on y Cajal fellowship RYC2021-033794-I and AC by the Severo Ochoa fellowship PA-23-BP22-019.

\appendix

\section{SO(6) spinors on $\mathbb{CP}^3$}\label{S7KSE}
In this section we give details on the derivation of the spinors in the \textbf{6} of SO(6) on $\mathbb{CP}^3$ that we make use of in section \ref{eq:derbilins}.

We start our derivation with the 7-sphere parametrised as a U(1) fibration over $\mathbb{CP}^3$ as
\begin{align}
ds^2(\text{S}^7)&= ds^2(\mathbb{CP}^3)+ (d\tau+ \eta)^2,~~~~4\eta=\cos^2\zeta(d\psi+ 2\eta_1)-\sin^2\zeta(d\psi+ 2\eta_2),\nn\\[2mm]
ds^2(\mathbb{CP}^3)&=d\zeta^2+\frac{1}{4}\cos^2\zeta ds^2(\text{S}^2_1)+\frac{1}{4}\sin^2\zeta ds^2(\text{S}^2_2)+\frac{1}{4}\cos^2\zeta\sin^2\zeta(d\psi+\eta_1+\eta_2)^2,\nn\\[2mm]
ds^2(\text{S}^2_i)&= d\theta_i^2+\sin^2\theta_id\phi_i^2,~~~\eta_i= \cos\theta_id\phi_i,
\end{align}
such that the SO(6) invariant Kahler form is given by
\beq
J_2=\frac{1}{2}\cos\zeta\sin\zeta d\zeta\wedge (d\psi+\eta_1+\eta_2)-\frac{1}{4}\cos^2\zeta\text{vol}(\text{S}_1^2)+\frac{1}{4}\sin^2\zeta\text{vol}(\text{S}_2^2).
\eeq
We elect the following vielbein basis on the 7-sphere
\begin{align}
e^1&=d\zeta,~~~e^2=\frac{1}{2}\cos\zeta d\theta_1,~~~e^3=\frac{1}{2}\cos\zeta \sin\theta_1d\phi_1,~~~e^4=\frac{1}{2}\sin\zeta d\theta_2,~~~e^5=\frac{1}{2}\sin\zeta \sin\theta_2\phi_2\nn\\[2mm]
e^6&= \frac{1}{2}\cos\zeta\sin\zeta(d\psi+\eta_1+\eta_2),~~~e^7= d\tau+\eta, \label{eq:frame}
\end{align}
and choose a basis of $d=7$ gamma matrices $\gamma_a$ (called $\gamma^{(7)}_a$ in the main text) as
\beq
\gamma_1=\sigma_1\otimes \mathbb{I}_2\otimes \mathbb{I}_2,~~~\gamma_{2,3,4}=\sigma_2\otimes \sigma_{1,2,3}\otimes \mathbb{I}_2,~~~\gamma_{5,6,7}=\sigma_3\otimes\mathbb{I}_2 \otimes \sigma_{1,2,3}
\eeq
for $\sigma_{1,2,3}$ the Pauli matrices,  which obey $i\gamma_{1...7}=\mathbb{I}$ and $B^{-1}\gamma_a B=-\gamma_a^*$ where $B$ ($B^{(7)}$ in the main text) is the intertwiner
\beq
B=\sigma_3\otimes \sigma_2\otimes \sigma_2
\eeq
and defines Majorana conjugation as $\xi^c=B\xi^{*}$ for $\xi$ a $d=7$ spinor. In terms of these, the Killing spinor equations on the 7-sphere
\beq
\nabla_a\xi_{\pm}=\pm\frac{i}{2}\gamma_a \xi_{\pm},
\eeq
are solved by
\begin{align}
\xi_{\pm}&= {\cal M}_{\pm}\xi^0_{\pm},\\[2mm]
{\cal M}_{\pm}&= e^{\frac{\zeta}{2}(\pm i\gamma_7+\gamma_{67})}e^{\frac{1}{8}(4\tau+\psi)(\pm\gamma_7-\gamma_{23})}e^{\frac{\theta_1}{4}(\pm i\gamma_2+\gamma_{37})}e^{\frac{\theta_1}{4}(\pm i\gamma_2+\gamma_{23})}e^{\frac{1}{8}(-4\tau+\psi)(\gamma_{16}-\gamma_{45})}e^{\frac{\theta_2}{4}(\gamma_{14}+\gamma_{56})}e^{\frac{\phi_2}{4}(\gamma_{16}+\gamma_{45})}\nn,
\end{align}
where $\xi^0_{\pm}$ are unconstrained constant spinors in $d=7$.

Each of $\xi_{\pm}$ contains 8 independent spinors transforming in the \textbf{8} of SO(8) which we can define as
\beq
\xi^{I}_{8\pm}={\cal M}_{\pm}\eta^I,~~~~\eta^I=\sum_{J=1}^8\delta^{IJ}.
\eeq
There are two ways that SO(8) can branch under an SO(6) subgroup outlined in \eqref{eq:branchingrules} so it seems natural that $\xi^{I}_{8\pm}$ should separately realise these branchings. 
It was argued in \cite{Legramandi:2020txf}, whose inventions regarding orientation we share, that $\xi^{I}_{8-}$ realises $\text{\textbf{8}} \to\text{\textbf{4}}_{-1}\oplus \overline{\text{\textbf{4}}}_{1}$ branching, and indeed it is quick to check that
\beq
\xi_{4-}^{\mathfrak{i}}=\frac{1}{\sqrt{2}}\left(\begin{array}{c}-i(\xi^{3}_{8-}+\xi^{8}_{8-})\\
-i(\xi^{2}_{8-}-\xi^{5}_{8-})\\
-(\xi^{1}_{8-}+\xi^{6}_{8-})\\
(\xi^{1}_{8-}-\xi^{6}_{8-})\end{array}\right)^{\mathfrak{i}},~~~~\xi_{4+}^{\mathfrak{i}}=\frac{1}{\sqrt{2}}\left(\begin{array}{c}i(\xi^{2}_{8-}+\xi^{5}_{8-})\\
i(\xi^{3}_{8-}-\xi^{8}_{8-})\\
(\xi^{4}_{8-}+\xi^{7}_{8-})\\
i(\xi^4_{8-}-\xi^7_{8-})\end{array}\right)^{\mathfrak{i}},
\eeq 
which require all components of $\xi^{I}_{8-}$ to span, are such that
\beq
\partial_{\tau}\xi_{4\pm}^{\mathfrak{i}}=\pm i \xi_{4\pm}^{\mathfrak{i}},~~~~\xi_{4-}^{\mathfrak{i}c}=\xi_{4+}^{\mathfrak{i}}
\eeq
consistent with this claim. Note that in the frame of \eqref{eq:frame} the spinoral Lie derivative ${\cal L}_{\partial_{\tau}}=\partial_{\tau}$, so the ordinary derivative is all that is needed to identify the U(1) charge under $\partial_{\tau}$. From these one can form an orthonormal set of Majorana spinors transforming in the $\text{\textbf{4}}\oplus \overline{\text{\textbf{4}}}$ of U(4) as
\beq
\xi^I_-= \frac{1}{\sqrt{2}}\left(\begin{array}{c}\xi_{4-}^{\mathfrak{i}}+\xi_{4+}^{\mathfrak{i}}\\
i(\xi_{4-}^{\mathfrak{i}}-\xi_{4+}^{\mathfrak{i}})\end{array}\right),
\eeq
however since these spinors are all charged under $\partial_{\tau}$ none of them survives the reduction to $\mathbb{CP}^3$. It is $\xi^{I}_{8+}$ that realises the $\text{\textbf{8}} \to\text{\textbf{6}}_0\oplus \text{\textbf{1}}_{-2}\oplus \text{\textbf{1}}_2$ branching under SO(6). Indeed it is a simple matter to confirm that 
\beq
\partial_{\tau}(\xi_{8+}^2-\xi_{8+}^5)=-2i (\xi_{8+}^2-\xi_{8+}^5) ~~~~\partial_{\tau}(\xi_{8+}^3-\xi_{8+}^8)=2i (\xi_{8+}^3-\xi_{8+}^8),
\eeq 
while the remaining 6 components of $\xi_{8+}^{I}$ independent of the above combinations do not depend on $\tau$ - it is only these that survive the reduction to $\mathbb{CP}^3$. The specific Majorana spinors of section \ref{eq:derbilins} transforming in respectively the \textbf{2} of SO(2) and \textbf{6} of SO(6) are
\begin{align}
\xi^{\mathfrak{a}}_2=\frac{1}{2}\left(\begin{array}{c}\xi^3_{8+}-\xi^8_{8+}+\xi^2_{8+}-\xi^5_{8+}\\
i(\xi^3_{8+}-\xi^3_{8+}-\xi^2_{8+}+\xi^5_{8+})\end{array}\right),~~~~\xi^{{\cal I}}_6=\frac{1}{\sqrt{2}}\left(\begin{array}{c}\xi_{8+}^1-\xi_{8+}^4\\
i(\xi_{8+}^1+\xi_{8+}^4)\\
\xi_{8+}^6-\xi_{7+}^1\\
i(\xi_{8+}^6+\xi_{7+}^1)\\
\frac{1}{\sqrt{2}}(\xi_{8+}^2+\xi_{8+}^5+\xi_{8+}^3+\xi_{8+}^8)\\
i\frac{1}{\sqrt{2}}(\xi_{8+}^2+\xi_{8+}^5-\xi_{8+}^3-\xi_{8+}^8)
\end{array}\right).
\end{align} 
This can be used to construct the explicit form of all the weak G$_2$ and SU(3)-structures of section \ref{eq:derbilins}, and more importantly the relations they obey.

\section{A no go for AdS$_2 \times \mathbb{CP}^3 \times \Sigma_2 $ in Type IIA}\label{sec:NogoTypeIIA}
In this section we rule out the possibility of AdS$_2 \times \mathbb{CP}^3 \times \Sigma_2$ in type IIA supergravity. 

Following the steps in section \ref{eq:derbilins} it is possible to show that the most general IIA bilinears consistent with \eqref{eq:noads3} are
\begin{align}
\psi_+ & =\frac{e^A}{16}\text{Re}\bigg(\psi^{(7)}_+ + \cos\beta \psi^{(7)}_- \wedge V \bigg),  ~~~~  \hat \psi_- = \frac{e^{A} \sin\beta}{16} \text{Re}\psi^{(7)}_- , \nn \\[2mm]
\psi^{(7)}_+ & = e^{i \alpha}e^{i e^{2C}\tilde{J}_2} + i e^{3C} e^{i \delta} \Omega_3 \wedge U, ~~~~ \psi^{(7)}_- = e^{i \alpha}e^{i e^{2C} \tilde{J}_2}\wedge U + i e^{3C} e^{i \delta} \Omega_3,\label{eq:IIAbilinears}
\end{align}
Using the bispinors \eqref{eq:IIAbilinears}
 and the supersymmetry constraints \eqref{BPS1}-\eqref{BPS3} we now prove that solutions preserving ${\cal N}=6$ supersymmetry on AdS$_2 \times \mathbb{CP}^3 \times \Sigma_2$ do not exist in Type IIA. We start by solving the constraints obtained by from \eqref{BPS2} and \eqref{BPS3}. The components along Re$\Omega_3$ of \eqref{BPS2} and $\tilde{J}_2 \wedge \tilde{J}_2 $ of \eqref{BPS3} give rise to the following zero form constraints
\begin{equation}
\sin \alpha = 0, \qquad m e^C = 6  e^A \sin \beta \sin \delta.
\end{equation}
These equations imply $\alpha = 0$ and $\sin \delta \neq 0$, otherwise we need $m=0$ which is Mink$_2$ not AdS$_2$ - note that $\sin \beta = 0$ is not compatible with AdS$_2$ in general \cite{Legramandi:2023fjr}. We turn our attention now to the following one-form and two-form constraints, we extract the following from the components along $J_2 \wedge J_2$  of \eqref{BPS2} and along Re$\Omega_3$ of \eqref{BPS3}
\begin{align}
& \cos \delta U \wedge V = 0, \\[2mm]
& d\left(e^{2A + 3C} e^{-\Phi} \sin \beta \cos \delta \right)+e^{A+3C} e^{-\Phi} m \cos \delta \cos \beta V + e^{A+2C} e^{-\Phi} (4 e^A \sin \beta - e^C m \sin \delta) U = 0. \nn
\end{align}
They cannot simultaneously be solved, leading  us to the non-existence of AdS$_2 \times \mathbb{CP}^3 \times \Sigma_2 $ solutions in Type IIA. There is a subtlety we must take into account though, in \cite{Macpherson:2023cbl} the authors presented a solution AdS$_3 \times \mathbb{CP}^3 \times I$ in Type IIA. AdS$_3$ admits a parametrisation as a foliation of AdS$_2$ over an interval which is of the form AdS$_2 \times \mathbb{CP}^3 \times \Sigma_2 $. That we do not find this solution is not a contradiction, it because we impose the ``no AdS$_3$ constraints'' \eqref{eq:noads3} before constructing \eqref{eq:IIAbilinears}.

\section{Extra details for the different classes of solutions}

In this Appendix we provide extra details for the different classes of solutions presented in the main text, that were omitted for the sake of clarity.

\subsection{Electric Page fluxes for the general class of solutions \eqref{NSNSIIBclass},\eqref{RRIIBclass}}\label{Sec:electricpagefluxes}

In this subsection we give the electric portion of the Page fluxes associated to the general class of solutions in the decomposition of \eqref{eq:pagefluxdecomp}:
\begin{align}
\hat{g}_1 & =\frac{c  m^3}{32 L^2} \left(d\left(\frac{\partial_2 h \left( x_1 \partial_1 h -h \right) \Box h }{\Delta_1} \right)- dx_1 x_1 \partial_2 \Box h + d x_2 \left(x_1 \partial_1 \Box h - \Box h \right)\right), \label{eq:genelpages}\\[2mm]
\hat{g}_3 & = c  m \bigg( d\left(\frac{\left(2 h -(x_2-l) \partial_2 h\right) \left(h - x_1 \partial_1 h \right) \Box h}{4 \Delta_1 }\right)\nn  \\[2mm]
& + \frac{1}{4}  \left( d x_2 (x_2 - l) \left(x_1 \partial_1 \Box h -\Box h \right) - d x_1 x_1 \left((x_2-l) \partial_2 \Box h - \Box h \right)\right)\bigg)  \wedge J_2,\nn\\[2mm]
 \hat{g}_5 & = \frac{c L^2}{m} \bigg[ -d\left(\frac{\left(h - x_1 \partial_1 h \right) \left((l-x_2)^2 \partial_2 h \Box h + 2 h \left(\partial_2 h + 2 (l-x_2) \Box h \right)\right)}{\Delta_1}\right) \nn \\[2mm]
 & -dx_1 x_1 \left(-2 (x_2-l) \Box h +(x_2-l)^2 \partial_2 \Box h +2 \partial_2 h \right)\nn\\[2mm]
 & +d x_2 \left((x_2 - l)^2 \left(x_1 \partial_1 \Box h -\Box h \right)-2 \left(h - x_1 \partial_1 h \right)\right) \bigg] \wedge J_2 \wedge J_2,\nn  \\[2mm]
 \hat{g}_7 & = \bigg[ -\frac{8  c  L^4 }{3 m^3} \bigg(d x_2 \left((x_2 - l)^3 \left(\Box h -x_1 \partial_1 \Box h  \right)-6 (x_2-l) x_1 \partial_1 h  +4 (x_2-l) h + 2 (x_2 - l) h \right)\nn \\[2mm]
 & + d x_1 \left(-3 (x_2-l)^2 x_1 \Box h +(x_2 - l)^3 x_1 \partial_2 \Box h+6 (x_2 - l) x_1 \partial_2 h - 6 x_1 h \right)\bigg)\nn \\[2mm]
 & + \frac{8}{3} \frac{c L^4 }{m^3}  d\left(\frac{\left(h-x_1 \partial_1 h \right) \left((x_2-l)^3 \left(-\partial_2 h \right) \Box h-6 (x_2-l) h \left( \partial_2 h -(x_2-l) \Box h \right)+4 h ^2 \right)}{\Delta_1}\right) \bigg] \nn \\[2mm]
 &  \wedge J_2 \wedge J_2 \wedge J_2.\nn
\end{align}
From these it is a simple matter to confirm that \eqref{eq: bianchielectric} does indeed hold.

\subsection{Electric Page fluxes for the solutions \eqref{andrea-class2},\eqref{andrea-class2RR}}\label{electricfluxsol}

In this subsection we give the relevant components of the electric Page fluxes of  the solutions \eqref{andrea-class2}, \eqref{andrea-class2RR}, corresponding to the choice $h=x_1 \hat{h}(x_2)$. Using the notation 
\beq
\hat F_-= \hat f_-+ \text{vol}(\text{AdS}_2)\wedge \hat g_-
\eeq
where $\text{vol}(\text{AdS}_2)\wedge \hat g_-$ is the electric portion of the Page flux polyform, we find
\begin{equation}
\begin{split}
\hat{g}_1 & = - \frac{1}{3} \frac{c}{2^5 L^2}  \hat{h}''' d(x_1^3), \\[2mm]
\hat{g}_3 & = \frac{1}{3} \frac{c}{2^2} \left(\hat{h}'' - \hat{h}''' (x_2 - l) \right) d(x_1^3) \wedge J_2. \\[2mm]
\end{split}
\end{equation}
There is also a non trivial $\hat{g}_5$ and $\hat{g}_7$, but these does not enter our analysis - they can be extracted from \eqref{eq:genelpages} after imposing the appropriate restriction on $h$.

\subsection{A second class of solutions for $\Sigma_2$ a strip}\label{sec:strip2}

A second class of solutions for the Riemann surface a strip can be obtained by choosing $h=x_2 \hat{h}(x_1)$. This class is given by,
\begin{equation}
\begin{split} \label{newsol}
\frac{ds^2}{L^2}&= \sqrt{\Delta_1}\bigg[\frac{x_2 \hat{h}}{\Delta_2}ds^2(\text{AdS}_2)+ \frac{1}{m^2}\bigg(\frac{8}{ x_2 \hat{h}''}ds^2(\mathbb{CP}^3)+ \frac{1}{ x_2 \hat{h}}(dx_i)^2\bigg)\bigg],~~~~e^{-\Phi}=\frac{m^3 x_2^2 \sqrt{\hat{h}} \hat{h}''^{3/2} }{16\sqrt{2}L^4\Delta_1},\nn\\[2mm]
H & = dB,~~~~\frac{B}{L^2}=\left(\frac{x_2^2 \hat{h} \hat{h}' }{\Delta_2}+x_1\right)\text{vol}(\text{AdS}_2)+ \frac{8}{m^2}\left(\frac{\hat{h}}{x_2 \hat{h}''}-(x_2-l)\right)J_2,
\end{split}
\end{equation}
where now
\begin{equation}
\Delta_1 = 2 x_2^2 \hat{h} \hat{h}'' - \hat{h}^2, \qquad \Delta_2 = x_2^2 \left(2 \hat{h} \hat{h}''-\hat{h}'^2\right)-\hat{h}^2.
\end{equation}
The magnetic Page fluxes take the form
\begin{align}
\hat f_1&=F_1= \frac{m^3}{32L^4}\bigg[\hat{h}'' dx_1- x_2 \hat{h}''' dx_2-d\left(\frac{x_2^2 \hat{h} \hat{h}' \hat{h}''}{\Delta_1}\right)\bigg],\nn\\[2mm]
\hat f_3 & =\frac{m}{4L^2}\bigg[- l \hat{h}'' dx_1-x_2(x_2-l)\hat{h}''' dx_2 + d\left(\frac{ x_2^2 (x_2 + l) \hat{h} \hat{h}' \hat{h}'' }{\Delta_1}\right)\bigg]\wedge J_2,\nn\\[2mm]
\hat f_5 &  =\frac{1}{m}\bigg[\left( 2 \hat{h} -(x_2-l)(x_2+l)\hat{h}''\right)dx_1-x_2 \left( 2 \hat{h}'+(x_2-l)^2 \hat{h}''' \right)dx_2\nn\\[2mm]
&+d\left(\frac{x_2^2 \hat{h} \hat{h}' \left( - 2 \hat{h} + (x_2 - l) (3 x_2 + l) \hat{h}''  \right) }{\Delta_1}\right)\bigg]\wedge J_2\wedge J_2,\nn\\[2mm]
\hat f_7 & = \frac{8L^2}{m^3}\bigg[\left(- 2 l \hat{h} - \frac{1}{3} (x_2-l)^2 (2x_2+l)\hat{h}''\right)dx_1 - 2 \bigg( x_2 (x_2-l) \hat{h}' +\frac{1}{6} x_2 (x_2-l)^3 \hat{h}''' \bigg) dx_2 \nn\\[2mm]
& + d\left(\frac{x_2^2 \hat{h} \hat{h}' \left(-(2 x_2 - 6l) \hat{h} + (x_2-l)^2(5x_2+l)\hat{h}'' \right)}{3\Delta_1}\right)\bigg]\wedge J_2\wedge J_2\wedge J_2.
\end{align}
Contrary to what happened with the two classes of solutions for $\Sigma_2$ an annulus, one can check that this and the previous class \eqref{andrea-class2},\eqref{andrea-class2RR}  are not related under S-duality in any limit.

\end{document}